\documentclass[11pt,a4paper]{article}
\usepackage{jheppub}
\usepackage{amsmath}
\usepackage{amssymb}
\usepackage{graphicx,caption}
\usepackage{subcaption}
\usepackage{float}
\usepackage{verbatim}
\usepackage{hyperref}							 %Hyperref an/aus
\usepackage[font=footnotesize]{caption}
\usepackage{slashed}
\usepackage{siunitx}
\usepackage{nicefrac}
\usepackage{tabularx}
\usepackage{tabulary}
\usepackage{pstricks}
\usepackage{color}
\usepackage{subfiles}
\usepackage{ stmaryrd }
\usepackage[labelfont=bf]{caption}
\usepackage{cleveref}
\usepackage{bbm}
\usepackage{amsmath,amsfonts,braket,slashed,amssymb,tikz,bm,psfrag,graphicx,color,dsfont}

\newcommand{\delmu}[0]{\partial_{\mu}}

\newcommand{\del}[0]{\partial}
\newcommand{\ga}[0]{\gamma}
\newcommand{\Ga}[0]{\Gamma}
\newcommand{\e}[0]{\epsilon}
\newcommand{\Li}[0]{\text{Li}}
\renewcommand{\d}[0]{\ensuremath{\operatorname{d}\!}}

\begin{document}
\begin{titlepage}
	
	\begin{flushright}
		\normalsize
		MITP-24-033\\ 
		\today
	\end{flushright}
	
	\vspace{1.0cm}
	\begin{center}
		\Large\bf\boldmath 
		Two-loop contributions of axion-like particles to electromagnetic and chromomagnetic form factors
	\end{center}
	
	\vspace{0.5cm}
	\begin{center}
		Matthias Neubert$^{a,b}$ and Marvin Schnubel$^c$\\
		\vspace{0.7cm} 
		{\sl ${}^a$PRISMA$^+$ Cluster of Excellence \& Mainz Institute for Theoretical Physics\\
		Johannes Gutenberg University, 55099 Mainz, Germany\\[3mm]
		${}^b$Department of Physics \& LEPP, Cornell University, Ithaca, NY 14853, U.S.A.
		${}^c$Department of Physics, Brookhaven National Laboratory, Upton, NY 11973, U.S.A.}
	\end{center}
	
	\vspace{0.8cm}
	\begin{abstract}
		
		Axions and axion-like particles emerge in many models for physics beyond the Standard Model. Thus, they have gained increasing research interest in both experimental and theoretical physics apart from their original proposition as a solution to the strong $CP$-problem. Among other aspects it has recently been shown that ALPs can potentially provide a solution to the long-lasting discrepancy between theory and experiment of the anomalous magnetic moment of the muon. Provided that the ALP has flavor-violating couplings to leptons, they can also mediate flavor-violating decays like $\mu\to e\gamma$. Both processes are mediated through related form factors that we compute to two-loop order. We further show numerical implications of our calculations and how they might affect constraints on ALP couplings derived from experiments.
		
	\end{abstract}
	
\end{titlepage}

\tableofcontents
%\newpage

	\section{Introduction}
	Axions and axion-like particles (to which we will commonly refer to as ALPs in this work) are well-motivated extensions of the Standard Model (SM). As an emergent pseudo Nambu-Goldstone boson from the spontaneous breaking of a global $U(1)$ symmetry, they are part of many new physics models. Initially, they were introduced as a solution to the strong CP-problem by Peccei and Quinn and others \cite{Peccei:1977hh, Peccei:1977ur, Weinberg:1977ma, Wilczek:1977pj}, though it was soon noticed that ALPs can have serious connections to other open SM questions, too.  One example is the question of the large hierarchy in fermion masses with the proposed solution of the Froggatt--Nielsen mechanism that introduces a new scalar field together with a high energy $U(1)$ symmetry under which the fermions are charged appropriately and which is spontaneously  broken at lower energies. Identifying this underlying $U(1)$ symmetry with the Peccei--Quinn symmetry, or equivalently, identifying the phase of the new scalar particle with an ALP, results in an ALP with couplings to all SM fermions with a possibly rich flavor structure \cite{Ema:2016ops, Calibbi:2016hwq, Alanne:2018fns}. Furthermore, especially light ALPs have been shown to be viable cold dark matter candidates \cite{Ringwald:2016yge} and be able to provide a stochastic gravitational wave background in the early universe \cite{Machado:2018nqk,Machado:2019xuc,Madge:2021abk}. Consequently, ALPs have gained strong research interest by theory and experimental initiatives likewise in recent years.
	
	The QCD axion naturally couples to gluons when it is supposed to provide a solution to the strong CP problem and in fact most concrete ALP models feature strong couplings of the ALP to gauge bosons. Whereas in KSVZ models the SM fermions are uncharged under the PQ symmetry \cite{Kim:1979if,Shifman:1979if}, additional ALP-fermion couplings are present in DSFZ models \cite{Dine:1981rt,Zhitnitsky:1980tq}. Additionally, the mass of the ALP and the photon coupling are inversely related to each other, resulting in a narrow band in the parameter space where the ALP is a solution to the strong CP problem \cite{Ringwald:2012hr}. In this work we take a model-independent approach and make no further assumptions on the ALP coupling structure. We also keep the ALP mass and its photon coupling as free parameters.
	
	In \cite{Bauer:2020jbp,Bauer:2021mvw,Chala:2020wvs} it was shown that an ALP coupling to any SM particle in the UV region will eventually generate couplings to all SM particles in the low-energy effective theory by subsequently evolving the operators from higher to lower scales through their renormalization group (RG) equations and, at the low scale, matching the effective theory to the Standard Model. Whenever a mass threshold is crossed in the evolution process, the respective particle is integrated out and and a new matching is performed. This procedure is repeated until at energies below $\mu_0\approx\SI{2}{\giga\eV}$ the theory is matched onto a chiral perturbation theory ($\chi$PT), where the ALP interacts directly with the hadrons instead. Most importantly, these running and matching effects generate effective quark flavor-changing couplings in the down sector, even if the underlying UV theory is flavor-blind or -conserving. This opens a huge variety of possibilities to probe UV ALP couplings with numerous flavor experiments. 
	
	Since the SM features the individual lepton numbers as accidental symmetries, evolution effects alone cannot generate ALP couplings that are lepton flavor-changing. In principle, loops containing flavor-oscillating neutrinos could give rise to flavor-changing processes also in the charged sector. However, they are suppressed by factors of $\Delta m_\nu^2/m_W^2\approx\SI{e-26}{}$, and hence are typically neglected. If such lepton flavor-violating are already present at tree-level in the UV region, in \cite{Bauer:2020jbp} it was shown that they do not receive corrections from the evolution procedure. 
	
	QCD axions are typically thought to be light, $m_\text{a}<\SI{1}{\eV}$. Hence, QCD axions and ALPs in that mass region are best probed by light shining through wall (LSW) experiments, stellar and astrophysical probes. Bounds coming from flavor experiments are most stringent in the mass range of $\sim\SI{0.1}{\mega\eV}$ to $\SI{10}{\giga\eV}$. For this work, we will therefore assume the ALP mass to lie in this mass region. If the ALP is heavier than $\sim\SI{10}{\giga\eV}$, best limits arise from searches for exotic decays of the Z and Higgs boson, as well as searches for direct production at colliders. We refer the interested reader to \cite{Davidson:1981zd,Davidson:1984ik,Peccei:1986pn,Krauss:1986wx,Geng:1988nc,Celis:2014iua,Alves:2017avw,DiLuzio:2017ogq,Choi:2017gpf,MartinCamalich:2020dfe,Gelmini:1982zz, Anselm:1985bp, Bauer:2017nlg,Bauer:2017ris,Bauer:2019gfk,Bauer:2020jbp,Bauer:2021mvw,Bauer:2021wjo} for detailed overview over studies of ALPs with both flavor-conserving and flavor-violating couplings.
	
	In a common normalization of the ALP Lagrangian, the ALP couplings to gauge bosons are written as $c_{VV}\frac{\alpha_V}{4\pi}\frac{a}{f}F_{\mu\nu}\tilde F^{\mu\nu}$ where $F^{\mu\nu}$ is the gauge boson field strength tensor and $\tilde F^{\mu\nu}=\frac{1}{2}\epsilon^{\mu\nu\alpha\beta}F_{\alpha\beta}$ its dual\footnote{We choose the convention $\epsilon^{0123}=1$.}. Pulling out a factor of $\alpha_V/(4\pi)$ of the Wilson coefficients ensures the scale independence of these couplings up to two-loop order. Consequently, a $\psi_1\to\psi_2\gamma$ one-loop Feynman diagram featuring an ALP-photon coupling is, na\"ively estimated, contributing at the same order as the two-loop diagram where the direct ALP-photon coupling is replaced by a fermion loop. The work at hand is therefore dedicated to study the impact of such two-loop effects thoroughly. Note that this is not only an academic exercise, but an important contribution to consistently study ALP effects in processes like the decay $\mu\to e\gamma$ and the anomalous magnetic moment of the muon  and the electron, $(g-2)_\mu$ and $(g-2)_e$.
	
	Two-loop ALP corrections to dipole moments have been calculated previously in the context of the anomalous magnetic moment of the muon in \cite{Buen-Abad:2021fwq}. In this work, we first re-evaluate the calculation in a different basis, eliminating the need for an additional subtraction to render the expressions finite. We then generalize the computation to flavor-changing currents. Additionally we transfer our findings to dipoles involving non-abelian fields and study the related contribution to the chomomagnetic moment of the top quark. 
	
	\section{ALPs coupling to the SM}
	\subsection{The effective ALP Lagrangian}
	The Lagrangian containing the ALP and its interactions with Standard Model particles is given by
	\begin{equation}
	\label{eq:LeffIR}
	\begin{aligned}
		\mathcal{L}_{\text{eff}}^{D\leq5}=&\frac12 \left(\del^\mu a\right)\left(\del_\mu a\right)-\frac{m_{a,0}^2}{2}a^2+\frac{\delmu a}{f}\sum_f \left(\bar{\psi}_L \bm{k}_F\gamma^\mu\psi_L+\bar{\psi}_R\bm{k}_f\gamma^\mu\psi_R\right)+c_{GG}\frac{\alpha_s}{4\pi}\frac{a}{f}G^a_{\nu\mu}\tilde G^{\mu\nu,a}\\
		&+c_{\gamma\gamma}\frac{\alpha}{4\pi}\frac{a}{f}F_{\mu\nu}\tilde F^{\mu\nu}+c_{\gamma Z}\frac{\alpha}{2\pi s_w c_w}\frac{a}{f}F_{\mu\nu}\tilde Z^{\mu\nu}+c_{ZZ}\frac{\alpha}{4\pi s_w^2 c_w^2}\frac{a}{f}Z_{\mu\nu}\tilde Z^{\mu\nu}\,,
	\end{aligned}
	\end{equation}
	where $G^a_{\mu\nu}$ and $F_{\mu\nu}$ are the gluon and photon field-strength tensors and $\tilde F^{\mu\nu}=\frac12\epsilon^{\mu\nu\alpha\beta}F_{\alpha\beta}$ and equivalently for $\tilde G$ are the dual tensors. The sine and cosine of the weak mixing angle are abbreviated to $\sin\theta_w\equiv s_w$ and $\cos\theta_w\equiv c_w$, respectively. The couplings are defined below the electroweak scale, however, we keep the top quark and $Z$ bosons as propagating degrees of freedom. Because of their high mass, the contributions of $Z$ bosons to the processes studied here are negligible. At the classical level the Lagrangian the ALP couplings to the SM fields are protected by an approximate shift symmetry $a\to a+\text{ constant}$. While the derivative couplings to fermions naturally feature this symmetry, the additional terms in the couplings to gauge bosons can be removed by field redefinitions. Due to instanton effects the coupling to gluons only respects a discrete version of the shift symmetry, realized as $a\to a+n\pi f/c_{GG}$, where $n$ is a natural number \cite{Weinberg:1977ma,Wilczek:1977pj}. The effective ALP mass $m_a$ is given as the sum of an explicitly shift-symmetry breaking ALP mass $m_{a,0}$ and the contribution of non-perturbative QCD dynamics and reads at lowest order in chiral perturbation theory \cite{Shifman:1979if,Bardeen:1978nq,DiVecchia:1980yfw}
	\begin{equation}
		\label{eq:ALPmass}
		m_a^2=m_{a,0}^2\left[1+\mathcal{O}\left(\frac{f_\pi^2}{f^2}\right)\right]+c_{GG}^2\frac{f_\pi^2m_\pi^2}{f^2}\frac{2m_u m_d}{(m_u+m_d)^2}\,,
	\end{equation}
	with $f_\pi\approxeq\SI{130}{\mega\eV}$ the pion decay constant. 
	
	We find that for the computation of $\psi_1\to\psi_2 V$ $(V=g,\gamma)$ form factors it is helpful to work with an alternative formulation of the Lagrangian in a different basis by applying the equations of motion to the SM fermions
	\begin{equation}
		\label{eq:LeffIRalt}
		\begin{aligned}
			\mathcal{L}_{\text{eff}}^{D\leq5}=&\frac12 \left(\del^\mu a\right)\left(\del_\mu a\right)-\frac{m_{a}^2}{2}+\mathcal{L}_\text{ferm}\\
			&+\tilde c_{GG}\frac{\alpha_s}{4\pi}\frac{a}{f}G^a_{\nu\mu}\tilde G^{\mu\nu,a}+\tilde c_{\gamma\gamma}\frac{\alpha}{4\pi}\frac{a}{f}F_{\mu\nu}\tilde F^{\mu\nu}\,.
		\end{aligned}
	\end{equation}
	Here,	
	\begin{equation}\label{eq:tilded}
		\tilde c_{\ga\ga}=c_{\ga\ga}+\sum_f N_c^f Q_f^2 c_{ff},\qquad\text{and}\qquad\tilde c_{GG}=c_{GG}+\frac12\sum\limits_q c_{qq}\,,
	\end{equation}
	where the sum of the first term runs over all SM fermions $f$ and in the second term it runs over all quark states $q$. The couplings $c_{ff}$ and $c_{qq}$ are related to the couplings in the Lagrangian \eqref{eq:LeffIR} via
	\begin{equation}
		\label{eq:cll}
		\begin{aligned}
			c_{f_if_i}&=[k_f]_{ii}-[k_F]_{ii}\,.\\
		\end{aligned}
	\end{equation}
	It is only in these combinations that the coupling parameters of the original Lagrangian can appear in physical observables \cite{Bauer:2020jbp,Chala:2020wvs}. The ALP-fermion Lagrangian is given by 
	\begin{equation}
		\label{eq:alpfermion}
		\begin{aligned}
			\mathcal{L}_\text{ferm}=-\frac{i a}{2 f}\sum_f\big[&(m_{f_i}-m_{f_j})[k_f+k_F]_{ij}\bar{f}_if_j\\
			&+(m_{f_i}+m_{f_j})[k_f-k_F]_{ij}\bar{f}_i\ga_5f_j\big]\,.
		\end{aligned}
	\end{equation}
	
	The suppression scale $f$ of the dimension-$5$ operators is related to the scale of global symmetry breaking by $\Lambda=4\pi f$. Often one decides to eliminate $f$ in favor of the axion decay constant $f_a$ under the relation $f_a\equiv-f/(2c_{GG})$. In the literature, this is often done when dealing with QCD axions. Note that the ALP-fermion couplings are suppressed with the fermion masses in this alternative formulation, allowing us to neglect couplings to neutrinos. 
	
\begin{figure}[t]
		\begin{center}
			\includegraphics[width=\textwidth]{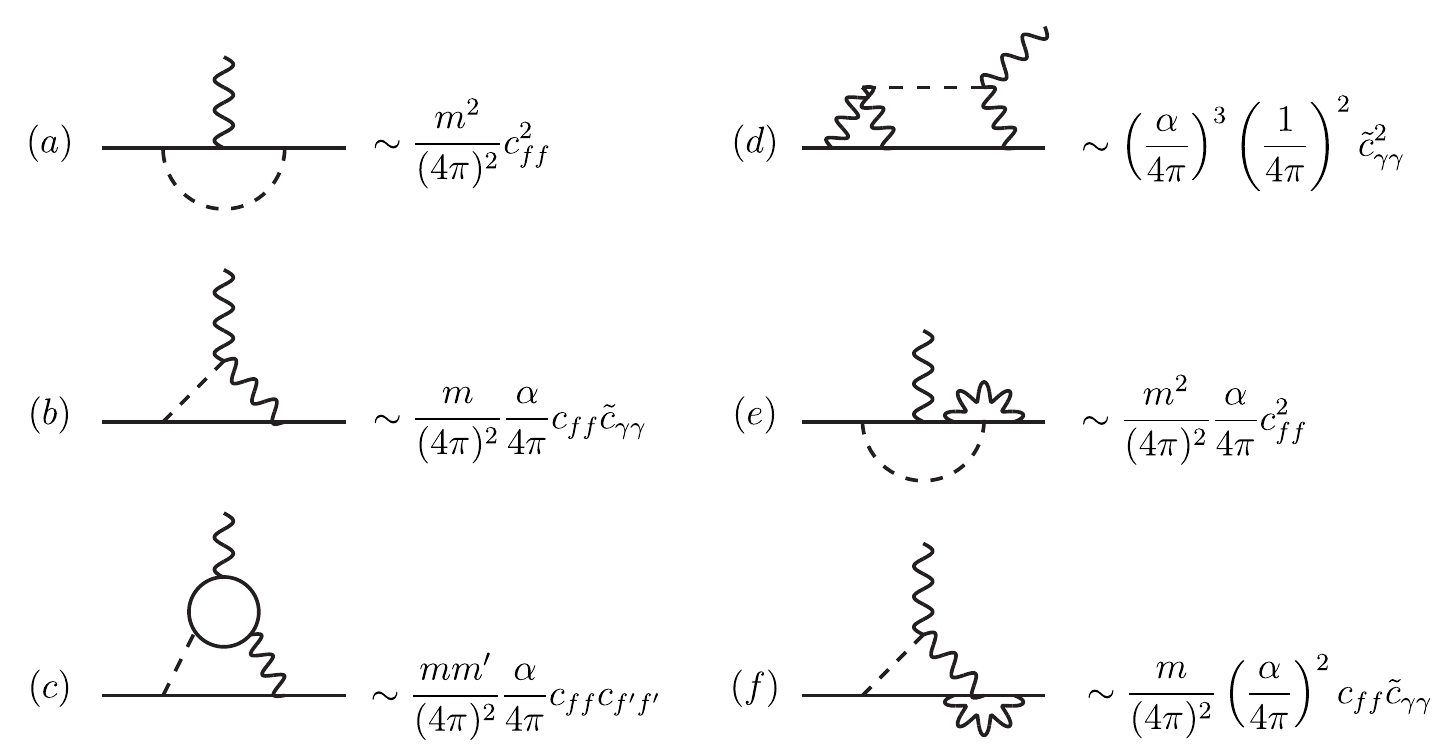}
			\caption{Representative diagrams contributing to ALP-induced $\psi\to\psi\gamma$ form factors up to two-loop order, excluding multiple ALP exchanges. Diagrams (a) and (b) are the one-loop diagrams. Diagrams (e) and (f) present loop-corrections to (a) and (b), respectively, and are therefore suppressed. Diagram (c) contributes at the same order in perturbation theory as diagram (b), even though it is a two-loop correction, and therefore its contribution must be taken into account for a consistent treatment if $c_{ff}\sim c_{\gamma\gamma}$ is assumed. In our choice of normalization of the ALP-gauge boson coupling diagram (d) is highly suppressed in perturbation theory. \label{fig:samemag}}
		\end{center}
	\end{figure}

	In the Lagrangians \eqref{eq:LeffIR} and \eqref{eq:LeffIRalt} we pulled out a normalization factor $\alpha_i/(4\pi)$ for the gauge boson couplings, as can be found in many explicit ALP models in the literature \cite{Bauer:2017ris,Bauer:2021mvw,Cornella:2019uxs,Chala:2020wvs}. This ensures that the ALP-gauge boson couplings $c_{VV}$ are scale independent at least up to two-loop order, and the scale dependence of the $\tilde c_{VV}$ couplings is fully governed by the evolution of the diagonal fermion couplings \cite{Chetyrkin:1998mw,Bauer:2020jbp,Chala:2020wvs}. As a consequence, we need to take certain two-loop graphs into account when studying $\psi_1\to\psi_2 V$ observables for the initial assumption that $c_{ff}\sim c_{VV}\sim\mathcal{O}(1)$, because they contribute with the same power of the QED/QCD coupling constant as other one-loop diagrams. This circumstance is illustrated in figure \ref{fig:samemag}. Here we present an estimate for the size of the contribution of representative classes of one and two-loop diagrams for the flavor conserving process $\psi_1\to\psi_1\ga$. We limit ourselves to one exchange of an ALP, as further exchanges are suppressed by additional factors of $1/f$. The diagrams (a) and (b) represent the two possible classes of one-loop diagrams. Diagrams (e) and (f) are loop corrections to the aforementioned cases and are therefore of subleading power in perturbation theory. This is not true for diagrams (c) and (d). From the tentative estimate one expects that inserting the fermion loop into the ALP-photon vertex in (c) contributes at the same order of perturbation theory as the direct contribution (b). For a consistent treatment of ALP induced $\psi_1\to\psi_2\gamma$ form factors this contribution must therefore be taken into account. It is similar to the well-known Barr-Zee diagrams \cite{Barr:1990vd} when the ALP is exchanged with a Higgs boson. Due to our choice of normalization the two-loop diagram with two ALP-photon vertices (d) is severely suppressed in perturbation theory. In models where the ALP-photon coupling is enhanced instead, they can give major contributions to the amplitude.

	\subsection{Form factors}
	In this section we will show how to derive the electromagnetic form factors 	relevant for $\psi_1\to\psi_2\ga$ processes. This can easily be generalized to the non-abelian case with the obvious replacements and insertions of color factors.
	
	The matrix element of a $\psi_1\to\psi_2\ga$ process can generally be written as
		\begin{equation}
		\mathcal{M}^\mu=\bar{\ell}_1(p_1)\Gamma^\mu\ell_2(p_2)\,,
		\label{eq:matrixelement}
	\end{equation}
	where $p_1$ ($p_2$) denotes the momentum of the initial- (final-) state particle. We furthermore define $p\equiv p_1+p_2$, and $q\equiv p_1-p_2$ is the momentum of the outgoing photon. If both initial- and final-state fermions are identical the matrix element can be parametrized in terms of four linearly independent form factors as follows
	\begin{align}\label{eq:ffactorsieqj}
		\hspace{-.3cm}\bar u_i (p_2)\Gamma^\mu(p_1,p_2)\,u_i(p_1) &= \bar u_i (p_2)\bigg[ F_2^{i\to i}(q^2)\big(p^\mu-2m_i\gamma^\mu\big)+2m_i F_3^{i\to i}(q^2)\gamma^\mu \notag \\
		&\phantom{=}+ F_2^{5,i\to i}(q^2)p^\mu \gamma_5+F_3^{5,i\to i}(q^2)\Big(q^\mu +\frac{q^2}{2m_i}\gamma^\mu\Big)\gamma_5\,
		\bigg]u_i(p_1)\,.
	\end{align}
	Note that two additional form factors $F_1^{(5)}$ that are proportional to $\gamma^\mu$ and $\gamma^\mu\gamma_5$, respectively, can be eliminated through application of the Ward identity $q_\mu\Gamma^\mu=0$. At tree-level in the SM equation \eqref{eq:ffactorsieqj} takes the form $\Gamma_{\text{SM},0}^\mu=Qe\gamma^\mu$, where $Q$ is the electric charge of the fermion. If the fermions are leptons of flavor $i$, one can read off the anomalous magnetic moment defined by $a_i=\frac{(g-2)_i}{2}$ as
	\begin{equation}\label{eq:anmomdef}
		a_i=\frac{2m_i}{e}F_2^{i\to i}(q^2=0).
	\end{equation}
	Furthermore, the electric dipole moment $d_i$ of a fermion with flavor $i$ is given by
	\begin{equation}\label{eq:edmdef}
		|d_i|=\frac{1}{2}\left| F_2^{5,i\to i}(q^2=0)\right|.
	\end{equation}

	In the case that initial and final states are of different fermion flavors instead, the parametrization reads
	\begin{align}\label{eq:ffactorsgeneral}
	\hspace{-.3cm}\bar u_j (p_2)&\Gamma^\mu(p_1,p_2)\,u_i(p_1) =\notag\\
	& \bar u_j (p_2)\bigg[ F_2^{i \to j}(q^2)\big(p^\mu-(m_i+m_j)\gamma^\mu\big)+F_3^{i \to j}(q^2)\Big(q^\mu -\frac{q^2}{m_i-m_j}\gamma^\mu\Big)\\
	&\phantom{=}+ F_2^{5,i \to j}(q^2)\big(p^\mu+(m_i-m_j)\gamma^\mu\big)\gamma_5+F_3^{5,i \to j}(q^2)\Big(q^\mu +\frac{q^2}{m_i+m_j}\gamma^\mu\Big)\gamma_5\,
	\bigg]u_i(p_1)\,.\notag
	\end{align}
	Note that in equations \eqref{eq:ffactorsieqj} and \eqref{eq:ffactorsgeneral} we keep the momentum $q$ of the photon general, even though it would be sufficient to take $q^2\equiv0$, i.e. the on-shell limit, for our purposes. The reason why we do so is that if the photon is off-shell it can give rise to secondary lepton pair production that could induce experimental signatures like $\mu^-\to e^-e^+e^-$.

	Flavor-changing neutral currents (FCNCs) are loop-suppressed in the Standard Model. Since the lepton numbers are individually conserved, lepton flavor-changing processes are even strictly forbidden. Hence, we assume for this work that the necessary flavor change stems from an ALP interaction instead. The branching ratio of such an ALP-induced flavor-changing decay is given by
	\begin{equation}
		\label{eq:branchingratiogeneral}
		\text{Br}(\psi_1\to\psi_2\gamma)=\frac{m_1^3}{8\pi\Gamma_1}\left(|F_2^{1\to2}(0)|^2+|F_2^{5, 1\to2}(0)|^2\right)\,,
	\end{equation}
	with $\Gamma_1$ the decay width of the initial state fermion. This formula is valid up to leading order in the expansion in $m_2^2/m_1^2$. The mass hierarchies between the different fermion generations ensure that this is indeed a sufficient approximation. 
	
	\section{Off-shell ALP-photon vertex}
	\begin{figure}[t]
		\begin{center}
			\includegraphics[width=0.6\textwidth]{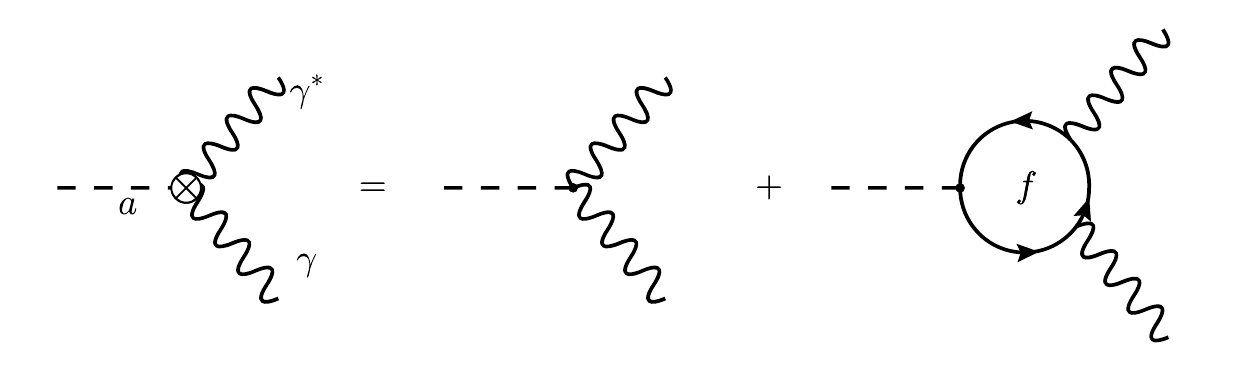}
			\caption{Feynman diagrams contributing to the effective ALP-photon vertex, including the one-loop correction from fermion loops. \label{fig:Agagavertex}}
		\end{center}
	\end{figure}
	As an intermediate step, we compute the fermion loop correction to the ALP-photon vertex, see figure \ref{fig:Agagavertex}. Here, we keep one of the photons as well as the ALP off-shell. The momentum of the ALP is denoted by $k$, and $p$ and $q$ represent the momenta of the off- and on-shell photons, respectively. The loop integral is finite and can be evaluated in $d=4$ spacetime dimensions. Including both tree-level and loop-induced diagrams, we obtain for the ALP-photon vertex
	\begin{align}
		\label{eq:alpgammavertex}
		\Ga^{\mu\alpha}_{a\ga\ga^\star}(p^2,k^2)=&\frac{i\alpha}{\pi f}\e^{\mu\alpha\beta\rho}q_\beta p_\rho\nonumber\\
		&\times\bigg\{c_{\ga\ga}+\sum_f N_c^f Q_f^2 c_{ff}\bigg[1-\int\limits_0^1\d x\int\limits_0^1\d y\frac{m_f^2}{m_f^2-x\bar{x}(yp^2+\bar{y}k^2)-i\e}\bigg]\bigg\}\,,
	\end{align}
	with $\bar{x}=1-x$ and the polarization indices of off- and on-shell photons denoted as $\alpha$ and $\mu$. Since $k=p+q$ we may replace $yp^2+\bar{y}k^2=k^2-2y\,k\cdot q=(k-y\,q)^2$ in the denominator. We find that the term involving $k\cdot q$ can be neglected when initial and final state lepton coincide, as in the diagrams contributing to $(g-2)_\mu$. This renders the $y$-integration trivial. Note that this is not true for diagrams contributing to flavor-changing processes. It can be observed that heavy fermions decouple from the vertex, whereas light fermions contribute $1$ inside the rectangular bracket. For the vertex with two on-shell photons a fermion is considered light when it fulfills $m_f^2\ll m_a^2$, in the case of one off-shell photon the relevant condition is $m_f^2\ll|k^2|$ instead. 
	
	We emphasize that the fermion-loop insertion in the effective ALP-$\ga\ga$ vertex in figure \ref{fig:2loop} also includes light quarks. If the momentum variables $|p^2|$ and $|k^2|$ both take small values, of order GeV$^2$ or less, these contributions are sensitive to hadronic effects and cannot be calculated reliably using perturbation theory. The internal three-point function connecting the two photons and the ALP should then be replaced by a non-perturbative correlator, which could be studied using lattice QCD. For simplicity, we ignore these subtleties in the following discussion. Note that our treatment is correct if either the ALP or the initial or final state fermion is much heavier than the QCD scale $\Lambda_\text{QCD}$. This means, in particular, that our analysis of the chromomagnetic moment of the top quark (section \ref{sec:chromotop}) does not receive non-perturbative corrections due to light quark loops.
	
	\section{General expressions for the form factors at two-loops}
	The one-loop diagrams contributing to the form factors have already been calculated in the literature \cite{Bauer:2019gfk,Bauer:2021mvw,Cornella:2019uxs}, which is why we focus on the two-loop graphs here. Moreover, we only consider diagrams with one ALP exchange, as graphs with further ALP exchanges are suppressed by additional factors of the large new physics scale $1/f$. 
	\begin{figure}[t]
		\begin{center}
			\includegraphics[width=0.35\textwidth]{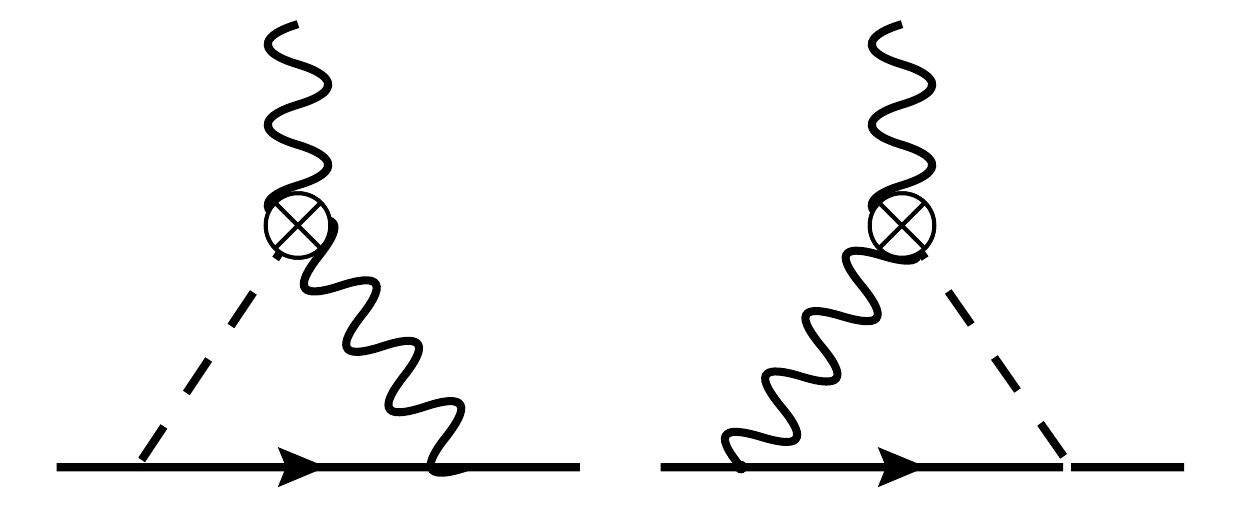}
			\caption{Feynman diagrams for $\psi_1\to\psi_2\gamma$ form factors with an effective ALP-photon coupling.\label{fig:2loop}}
		\end{center}
	\end{figure}
	
	We insert the vertex function \eqref{eq:alpgammavertex} into the diagrams, giving rise to the graphs shown in figure \ref{fig:2loop}. Since the off-shell vertex falls off like $1/k^2$ for large loop momenta (see \eqref{eq:alpgammavertex}), the graphs with internal fermion loops are UV finite and can be evaluated in $d=4$ spacetime dimensions. For the processes we are interested in, only the form factors $F_2^{(5)}(q^2)$ yield a contribution. In the relevant limit of $q^2=0$, i.e. an on-shell photon in the final state, we obtain
	\begin{equation}
		\label{eq:formfactortotal}
		F_2^{(5)}(q^2=0)=-\frac{e\alpha q}{16\pi^3}(m_1\pm m_2)\frac{|k_f|_{12}\mp |k_F|_{12}}{f}\bigg(\frac{\tilde c_{\ga\ga}}{f}\mathcal{I}_1+\sum\limits_f N_c^f Q_f^2\frac{c_{ff}}{f}\mathcal{I}_2\bigg)\,,
	\end{equation}
	where $q$ is the charge of the initial and final state fermion, and $Q_f$ and $N_c^f$ are the charge and number of flavors of the fermions in the one-loop correction in the effective ALP-photon vertex. $\mathcal{I}_1$ and $\mathcal{I}_2$ are the one and two-loop parameter integrals, respectively, and read
	\begin{equation}
		\label{eq:I12def}
		\begin{aligned}
			\mathcal{I}_1=&\int\limits_0^1\d x\int\limits_0^x\d y\bigg(2\delta_2+\ln\left(\frac{\mu^2}{\Delta_1}\right)+\ln\left(\frac{\mu^2}{\Delta_2}\right)\\
			&\qquad+\frac{m_1+m_2}{2}(x-y)\left[\frac{m_1(-1+x)+m_2(1-y)}{\Delta_1}+(m_1\leftrightarrow m_2)\right]\bigg)\,,\\
			\mathcal{I}_2=&\frac{1}{2}\int\limits_0^1\!\!\d x\frac{m_f^2}{x\bar{x}}\int\limits_0^1\!\!\!\d y\left[4(A+A')-(m_1+m_2)(m_1B+m_2B')+(m_1^2-m_2^2)(a+a'-C-C')\right]
		\end{aligned}
	\end{equation}
	with 
	\begin{equation}
		\label{eq:deltadef}
		\begin{aligned}
			\Delta_1=&m_a^2(1-x)+m_1^2(-1+x)(x-y)+m_2^2(1-y)(x-y)\,,\\
			\Delta_2=&m_a^2(1-x)+m_1^2(1-y)(x-y)+m_2^2(-1+x)(x-y)\,.
		\end{aligned}
	\end{equation}
	In \eqref{eq:I12def} the functions $a$, $A$, $B$ and $C$ and their primed counterparts are Passarino-Veltman coefficient functions. They are obtained through the relations (with $\bar{m}^2=m_f^2/(x\bar{x})$)
	\begin{equation}
		\label{eq:PVdecom}
		\begin{aligned}
		&\int\frac{\d^4k}{(2\pi)^4}\frac{k^\mu}{(k^2-m_a^2)(k-q)^2((k-yq)^2-\bar{m}^2)((k-p_1)^2-m_2^2)}=\frac{1}{(4\pi)^2}\left(ap_1^\mu+bq^\mu\right)\,,\\
		&\int\frac{\d^4k}{(2\pi)^4}\frac{k^\mu k^\nu}{(k^2-m_a^2)(k-q)^2((k-yq)^2-\bar{m}^2)((k-p_1)^2-m_2^2)}\\
		&\hspace{5cm}=\frac{1}{(4\pi)^2}\left(Ag^{\mu\nu}+Bp_1^\mu p_1^\nu+C(p_1^\mu q^\nu+q^\mu p_1^\nu)+Dq^\mu q^\nu\right)\,.
		\end{aligned}
	\end{equation}
	The primed coefficients follow equivalently with the replacement $(p_1,m_2)\to(p_2,m_1)$. Moreover, $\delta_2=-3$ is a scheme-dependent constant derived from treating the Levi-Civita-tensor as a $d$-dimensional object. If it is seen as a $d=4$ dimensional object instead, $\delta_2=0$ \cite{Bauer:2017ris}. Note that the one-loop graphs are divergent and we choose to regularize them with dimensional regularization. The dependence on the regularization scale $\mu$ as well as the dependence on the scheme-dependent constant $\delta_2$ are eventually removed by adding a contribution to the form factor $F_2^{(5)}(q^2)$ from dimension-$6$ LEFT operators, as explained in~\cite{Galda:2023qjx}.
	
	While it is possible to obtain analytic expressions for $\mathcal{I}_1$ and reduce $\mathcal{I}_2$ to one unevaluated parameter integral, the equations are too long and unhandy to present them in this article in a convenient form. In the phenomenologically relevant cases, $m_1=m_2$ and $m_2\ll m_1$, they can be cast into more compact representations, however
	\begin{equation}
		\label{eq:I1limits}
		\begin{aligned}
			\mathcal{I}_1^{m_1=m_2}=&\delta_2+\ln\left(\frac{\mu^2}{m_1^2}\right)+3-h_2\left(\frac{m_a^2}{m_1^2}\right)\,,\\
			\mathcal{I}_1^{m_2\ll m_1}=&\delta_2+\ln\left(\frac{\mu^2}{m_1^2}\right)+2+g_2\left(\frac{m_a^2}{m_1^2}\right)
		\end{aligned}
	\end{equation}
	with
	\begin{equation}
		\label{eq:ghdef}
		\begin{aligned}
			h_2(x)=&1-\frac{x}{3}+\frac{x^2}{6}\ln x+\frac{x+2}{3}\sqrt{(4-x)x}\arccos\frac{\sqrt{x}}{2}\,,\\
			g_2(x)=&-\frac{x^2\ln x}{2(x-1)}+\frac12(x-1)\ln(x-1)\,.
		\end{aligned}
	\end{equation}
	The two-loop terms simplify to	
	\begin{equation}
		\label{eq:I2limits}
		\begin{aligned}
			\mathcal{I}_2^{m_1=m_2}=&\int\limits_0^1\d xF\left(y_x,r_1\right)\,,\\
			\mathcal{I}_2^{m_2\ll m_1}=&\int\limits_0^1\!\frac{\d x}{x\bar{x}}\int\limits_0^1\!\d u\bigg\{(1\!-\! u)ur_1\bigg[-(r_a-r_1u)x\bar{x}\ln\left(\frac{(r_a-r_1u)x\bar{x}}{1-r_1 ux\bar{x}}\right)-\ln(1-r_1 ux\bar{x})\bigg]\\
			&+3(1\!-\! u)(1\!-\!(r_a-r_1u)x\bar{x})\left[\Li_2\left(\frac{-1+(r_a-r_1u)x\bar{x}}{(r_a-r_1u)x\bar{x}}\right)\!-\!\Li_2\left(\frac{-1+r_ax\bar{x}}{(r_a-r_1u)x\bar{x}}\right)\right]\\
			&+\int\limits_0^{1}\!\d y\int\limits_0^{1-u}\!\d v\int\limits_0^{1-u-v}\!\d w\frac{r_a(1-u-v-w)+\frac{w}{x\bar{x}}+\frac32 r_1u(1-v-w y)}{\left(r_a(1-u-v-w)+\frac{w}{x\bar{x}}+\frac32 r_1u(1-v-w y)\right)^2}\bigg\}\,.
		\end{aligned}
	\end{equation}
Here, we introduced the abbreviations $r_a=m_a^2/m_f^2$ and $r_1=m_1^2/m_f^2$. The parameter integrals in the last row can all be computed analytically. However, their expressions are unwieldy long, which is why we decided to omit them here. In the first line
\begin{equation}
		\label{eq:yFdef}
		F(y_x,r_1)=\frac{1}{1-y_x}\left[h_2\left(\frac{r_a}{r_1}\right)-h_2\left(\frac{1}{x\bar{x}r_1}\right)\right]\,,\quad\text{and}\quad y_x=x(1-x)r_a\,.
	\end{equation}
The function $\mathcal{I}_2^{m_1=m_2}$ in the equal mass limit has first been computed in \cite{Buen-Abad:2021fwq}. In this calculation the loop function in the heavy mass limit needed to be subtracted to obtain the correct result. The reason why a similar procedure is not necessary in our calculation is that we changed the basis by applying the equations of motions, and as a result the ALP features no derivative couplings in this alternative form of the Lagrangian \eqref{eq:LeffIRalt}. The advantage of this alternative formulation becomes clear after a closer inspection of the effective ALP-photon vertex \eqref{eq:alpgammavertex}. In the infinite mass limit $m_f^2\gg k^2$ the rectangular bracket including the two-loop contribution vanishes. In the opposite limit $m^2\ll k^2$ the additional propagator renders the two-loop contribution finite, allowing us to perform the calculation in $d=4$ spacetime dimensions regardless of the mass of the inner fermion. However, we want to emphasize here that both basis choices are of course equivalent, and consequently our results fully agree with those of Ref. \cite{Buen-Abad:2021fwq}.
	
The remaining parameter integrals can be readily integrated numerically. In certain limits we find it interesting to report some useful explicit results. The limits $m_a^2\gg m_1^2$ and $m_f^2\gg m_{a,1}^2$ yield the same functional behavior for both the flavor-conserving and flavor-changing case.
\paragraph{Limit \boldmath$m_a^2\gg m_1^2$}
In the limit that the ALP is much heavier than the initial state fermion, one can take the limit $r=\frac{r_a}{r_1}\to\infty$ to obtain
\begin{equation}
	\label{eq:integrallimit1}
	\begin{aligned}
f(r_a)&=\lim\limits_{r\to\infty}\int\limits_0^1\d x\,F(y_x,r_1)\\
		&=\frac{-4}{\sqrt{r_a(r_a-4)}}\left\{\frac{\pi^2}{12}+\ln^2\left(\frac12\left(\sqrt{r_a}-\sqrt{r_a-4}\right)\right)+\Li_2\left[-\frac14\left(\sqrt{r_a}-\sqrt{r_a-4}\right)^2\right]\right\}\,,
	\end{aligned}
\end{equation}
Its asymptotic behavior is such that
\begin{equation}
	\label{eq:asympf}
	\begin{aligned}
		f(r_a)&=\ln\frac{m_f^2}{m_a^2}-2+\mathcal{O}\left(\frac{m_a^2}{m_f^2}\right);&m_f^2\gg m_a^2\,,\\
		f(r_a)&=-\frac{m_f^2}{m_a^2}\left[\ln^2\left(\frac{m_a^2}{m_f^2}\right)+\frac{\pi^2}{3}\right]+\mathcal{O}\left(\frac{m_f^4}{m_a^4}\right)\;;&m_f^2\ll m_a^2\,.
	\end{aligned}
\end{equation}
The fact that a light internal fermion decouples in the Barr-Zee graphs is more general and holds also if $m_a\sim m_1$. It follows that for an light internal fermion $(m_f^2\ll m_a^2)$ the sum of all contributions proportional to $c_{ff}$ is
\begin{equation}
	\label{eq:sumoflight}
	N_c^fQ_f^2c_{ff}\left[\ln\frac{\mu^2}{m_1^2}+\delta_2+3-h_2\left(\frac{m_a^2}{m_1^2}\right)+\mathcal{O}\left(\frac{m_f^2}{m_a^2}\right)\right]\,.
\end{equation}
In essence this means that virtual loop momenta are cut off for values below the scale $\text{max}(m_a,m_1)$.
\paragraph{Limit \boldmath$m_f^2\gg m_{a,1}^2$}
In the limit that the inner fermion is much heavier than both the initial state fermion and the ALP, one finds that
\begin{equation}
	\label{eq:integrallimit2}
	\lim\limits_{r_1,r_a\to0}\int\limits_0^1\d x\,F(y_x,r_1)=-\ln\frac{m_f^2}{m_1^2}+h_2\left(\frac{m_a^2}{m_1^2}\right)-\frac72+\mathcal{O}\left(\frac{m_{a,1}^2}{m_f^2}\right)\,.
\end{equation}
It follows that for an internal heavy fermion $(m_f^2\gg m_a^2)$ the sum of all contributions proportional to $c_{ff}$ is
\begin{equation}
	\label{eq:sumofheavy}
	N_c^fQ_f^2c_{ff}\left[\ln\frac{\mu^2}{m_f^2}+\delta_2-\frac12+\mathcal{O}\left(\frac{m_{a,1}^2}{m_f^2}\right)\right]\,.
\end{equation}
In essence, the virtual loop momenta are cut off for values below the scale $m_f$. For scales $\mu\ll m_f$ the heavy fermion decouples and can be integrated out.
\paragraph{Limit \boldmath$m_1^2=m_2^2=m_f^2\gg m_a^2$}
The limit where the initial and final state fermion coincide with the inner loop-fermion and the ALP is very light light is mainly relevant in the case of the chromomagnetic moment of the top quark. The integral in the first line of \eqref{eq:I2limits} then reads
\begin{equation}
	\label{eq:limitallesschwer}
	\lim\limits_{r_1=1,r_a\to0}\int\limits_0^1\d x\,F\left(y_x,r_1\right)=\frac23+\frac{2\pi^2}{9}-\frac{2\pi}{3}\frac{m_a}{m_f}+\left(\frac13+\frac{\pi^2}{9}\right)\frac{m_a^2}{m_f^2}+\mathcal{O}\left(\frac{m_a^3}{m_f^3}\right)\,.
\end{equation}
Note that for this special case the expansion is linear in $m_a/m_f$, giving that the ALP mass effects are not as strongly suppressed as in the other limits that we have discussed here. The opposite case that $m_1^2=m_2^2=m_f^2\ll m_a^2$ is already covered with the second equation in \eqref{eq:asympf}.

\subsection{Two-loop contributions proportional to \boldmath$c_{\ga\ga}^2$}
\label{sec:gagasq}
\begin{figure}[t]
	\begin{center}
		\includegraphics[width=\textwidth]{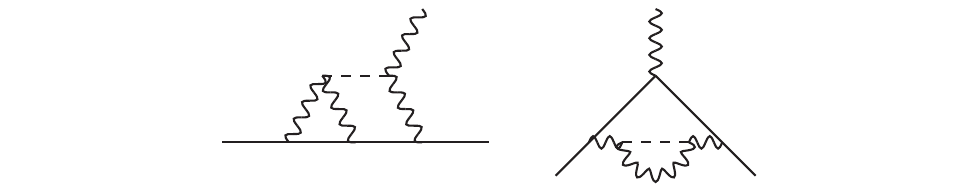}
		\caption{Feynman diagrams for the 2-loop contribution to the flavor-conserving process proportional to the ALP-photon coupling only. Left: Light-by-light scattering. Right: Photon vacuum polarization.\label{fig:2loopgaga}}
	\end{center}
\end{figure}

Two-loop diagrams involving two ALP-photon vertices yield amplitudes proportional to $c_{\ga\ga}^2$ and are highly suppressed in our normalization of the Lagrangian, see for example the scaling of diagram (d) in figure \ref{fig:samemag}. In models where the ALP-photon coupling is enhanced instead, they can give substantial contributions. The relevant diagrams consist of light-by-light (LbL) scattering diagrams and ALP insertions into the internal photon propagator that are similar to the corresponding contribution of pions to the SM vacuum polarization. The diagrams are shown in figure \ref{fig:2loopgaga}. We find it instructive to to give an estimate of their size and compare their effects with the two-loop diagrams described in the previous section, since they arise at the same loop order. In \cite{Marciano:2016yhf} the contributing diagrams have been calculated to leading logarithmic order. It was shown that while the LbL contribution gives rise to double logarithmic corrections, the vacuum polarization contribution only yields single logarithmic behavior and can therefore often be neglected. In \cite{Galda:2023qjx} the contributions where derived more thoroughly through the means of solving the renormalization group equations. Furthermore it was noticed that in the previous work the single logarithmic contribution was underestimated, yet the conclusion that the double logarithmic part gives the dominant contribution was correct. Translating their findings to the quantities used in this work, the contributions proportional to $c_{\ga\ga}^2$ read
\begin{equation}
	F_2^{c_{\ga\ga}^2}(q^2=0)=\frac{em_\mu^3}{(4\pi)^2}\left(\frac{\alpha}{\pi}\right)^2\left(\frac{c_{\ga\ga}}{f}\right)^2\frac38\left[\ln^2\frac{\mu^2}{m_\mu^2}+\left(2\delta_2+\frac{56}{9}-2h_2\left(x_\mu\right)+\frac{x_\mu}{3}\right)\ln\frac{\mu^2}{m_\mu^2}\right]\,,
\end{equation}
with $x_\mu=m_a^2/m_\mu^2$. The scale $\mu$ should be chosen as $\mu\sim\text{max}(m_\mu,m_a)$. However, if the ALP is too heavy it should be integrated out from the low-energy effective theory instead. Above the electroweak (EW) scale contributions from $Z$ bosons need to be taken into account as well. In our model where we all $Z$ bosons as propagating degrees of freedom also below the EW scale, their contributions are negligible.

The diagrams in figure \ref{fig:2loopgaga} are very similar to SM diagrams with LbL scattering with neutral pions and pion insertions in the photon propagator. They have been studied intensely in the literature as they are an important ingredient of the SM content supplying to the anomalous magnetic moment of the muon $a_\mu$. Note that diagrams of this type cannot give contributions to flavor-changing observables since the flavor-change must be provided directly by the ALP.
	
\section{Application to \boldmath$(g-2)_\mu$ and $\mu\to e\gamma$}
In this section we apply our findings to the phenomenological studies of the anomalous magnetic moment of the muon $a_\mu$ and the lepton flavor-violating decay $\mu\to e\gamma$. The anomalous magnetic moment defined via $a_\mu=(g-2)_\mu/2$ is one of the most precisely measured quantities in particle physics, measured in the Brookhaven~\cite{Muong-2:2006rrc} and Fermilab~\cite{Muong-2:2021ojo} experiments. However, it deviates from its predicted value in the SM. How large this deviation is cannot be stated without ambiguity. The reason for this is that there are currently two methods used for determining the hadronic vacuum polarization (HVP), the part which is currently giving the dominating contribution to the theory uncertainty. In the data-driven approach the HVP is extracted from $e^+e^-$-scattering data, resulting in a discrepancy between the measurement of $a_\mu$ and the prediction of over $5\sigma$ \cite{Aoyama:2020ynm}\footnote{The theory initiative paper derives their result based on various contributions to the SM prediction~\cite{Davier:2010nc,Aoyama:2012wk,Aoyama:2019ryr,Czarnecki:2002nt,Gnendiger:2013pva,Davier:2017zfy,Keshavarzi:2018mgv,Colangelo:2018mtw,Hoferichter:2019gzf,Davier:2019can,Keshavarzi:2019abf,Kurz:2014wya,Melnikov:2003xd,Masjuan:2017tvw,Colangelo:2017fiz,Hoferichter:2018kwz,Gerardin:2019vio,Bijnens:2019ghy,Colangelo:2019uex,Blum:2019ugy,Colangelo:2014qya}.}. Alternatively, it can be obtained using only input from lattice calculations by the Budapest, Marseille and Wuppertal (BMW) collaboration (BMW)~\cite{Borsanyi:2020mff}, yielding a better agreement with the experimental result. Since the lepton numbers are individually conserved in the SM, lepton flavor-violating (LFV) decays like $\mu\to e\gamma$ are forbidden. Currently, the best constraint on the respective muon branching ratio was obtained by the MEG collaboration and reads Br$(\mu\to e\ga)<\SI{4.2e-13}{}$ \cite{TheMEG:2016wtm}. Taking neutrino oscillations into account the SM expectation is $\text{Br}(\mu\to e\ga)|_\text{SM}\approx \SI{e-52}{}$. Therefore LFV observables present an excellent probe of new physics.

\subsection[Application to $a_\mu$]{Application to \boldmath$a_\mu$}\label{sec:amm}
An ALP coupling to SM particles gives rise to several contributions to $a_\mu$. At the one-loop level, they have been calculated in \cite{Bauer:2019gfk,Bauer:2017ris}. Here, we extend this computation by including the two-loop graph with an inserted fermion-loop. The final result is given by
	\begin{equation}
	\label{eq:deltaamu}
	\delta a_\mu=\!\!\frac{m_\mu^2}{f^2}\bigg(K_{a_\mu}(f,\mu)-\frac{c_{\mu\mu}^2}{16\pi^2}h_1\left(\frac{m_a^2}{m_\mu^2}\right)-\frac{\alpha}{8\pi^3}c_{\mu\mu}\underbrace{\bigg[\tilde c_{\ga\ga}\mathcal{I}_1^{m_1=m_2}+\sum\limits_f N_c^f Q_f^2c_{ff}\mathcal{I}_2^{m_1=m_2}\bigg]}_{=c_{\ga\ga}^\text{eff}}\bigg),
\end{equation}
and we have ignored the numerically sub-dominant contributions from the $Z$-boson. The parameter function $h_1$ was first calculated in \cite{Bauer:2017ris} and reads
\begin{equation}
	\label{eq:h1def}
	h_1(x)=1+2x+(1-x)x\ln x-2x(3-x)\sqrt{\frac{x}{4-x}}\arccos\frac{\sqrt{x}}{2}\,.
\end{equation}

\begin{figure}
	\begin{subfigure}[t]{0.5\textwidth}
		\centering
		\includegraphics[width=\textwidth]{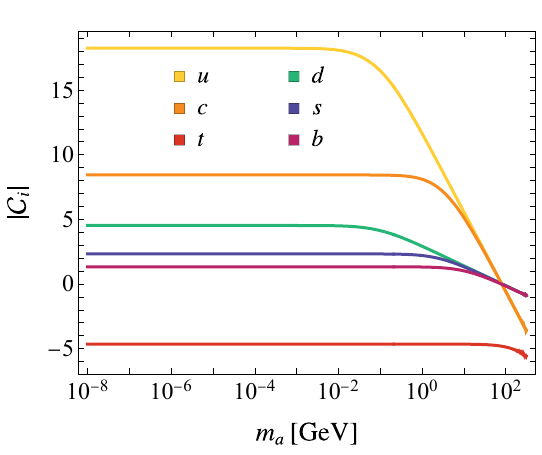}
	\end{subfigure}
	\begin{subfigure}[t]{0.5\textwidth}
		\centering
		\includegraphics[width=\textwidth]{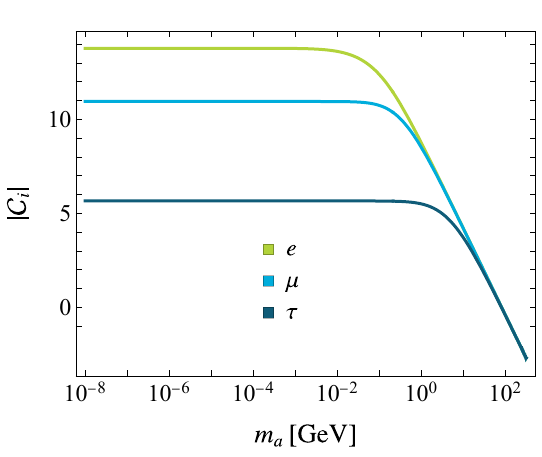}
	\end{subfigure}
	\caption{Coefficients $\mathcal{C}_i$ in \eqref{eq:cgagaeffdecompose} for quarks (left) and leptons (right). When the ALP is light when compared with the fermion in the loop, the effective coupling tends to constant plateau, and the value is entirely governed by the mass of the loop fermion. In the case that the ALP is the heaviest particle the effective coupling scales as $\sim N_c^fQ_f^2\ln \mu^2/m_a^2$. Hence the behavior is dictated by the ALP mass and the slope depends on the fermion type, i.e. up or down-type quark or lepton.\label{fig:fermionloops}}
\end{figure}

To disentangle the effects of the different fermion species in the loop, we decompose the effective ALP-photon coupling $c_{\ga\ga}^\text{eff}$ defined in \eqref{eq:deltaamu} into
\begin{equation}
	\label{eq:cgagaeffdecompose}
	c_{\ga\ga}^\text{eff}=c_{\ga\ga}\mathcal{I}_1^{m_1=m_2}+\sum\limits_i \mathcal{C}_i c_{ii}\,,
\end{equation}
where $i$ runs over all SM fermion states. In figure \ref{fig:fermionloops} we show the coefficients $\mathcal{C}_i$ for the contributions of the individual fermions for quarks (left) and leptons (right). The shape of the functions can be explained as follows: If the ALP is very heavy, the light fermion decouples and the contribution to the effective photon coupling is only dependent on the ALP mass, as dictated by equation \eqref{eq:sumoflight}. Essentially the coupling scales as $\lim\limits_{m_a\gg m_f}c_{\ga\ga}^\text{eff}\sim N_c^fQ_f^2\ln \mu^2/m_a^2$, explaining the linear pattern in the logarithmically scaled plot as well as the different slopes for up and down-type quarks, and leptons, respectively, based on the number of colors. In the opposite regime, $m_a^2\ll m_f^2$, the behavior is instead governed by \eqref{eq:sumofheavy}. The effective coupling reaches a plateau whose value only depends on the mass of the loop fermion. Note further that, in the up and down quark sector respectively, light quarks give a bigger contribution than heavy quarks do, as long as the quark can be considered light when compared with the ALP. This is expected, since in this regime essentially $\mathcal{C}_i\sim Q_{f_i}^2 \ln\mu^2/m_{f_i}^2$. To get a feeling how strong these contributions are, we compare the value of the plateau for $m_a^2\ll m_f^2$ with the value of the coefficient of $c_{\ga\ga}$ in that limit. We find $\lim\limits_{m_a\to0}\mathcal{I}_1^{m_1=m_2}=13.8$, where we have chosen the renormalization scale as the top quark mass $\mu=m_t$. Hence, especially light fermions give rise to sizable contributions to the effective ALP-photon coupling if one takes all ALP couplings to be of the same order $c_{\ga\ga}\sim c_{ff}$.

\begin{figure}[t]
	\begin{center}
		\includegraphics[width=\textwidth]{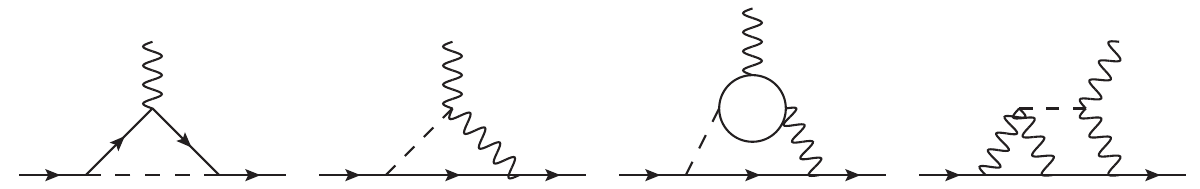}
		\caption{Leading order diagrams involving an ALP that contribute to the anomalous magnetic moment of the muon up to two-loop level. The contributions are proportional to $c_{\mu\mu}^2, c_{\mu\mu}\tilde c_{\ga\ga}, c_{\mu\mu}c_{ff}, c_{\ga\ga}^2$, reading from left to right. Note that for second and third diagram there exists a mirror diagram, where the ALP leg is interchanged with the photon leg. In the third diagram the inner loop fermion can run both clockwise and counter-clockwise. The fourth diagram has a mirror diagram where the ALP-photon vertex with one on-shell photon is exchanged with the other, additionally a diagram where the two rightmost legs are interchanged and the mirror of that one, too.  \label{fig:compamu}}
	\end{center}
\end{figure}

\begin{figure}[t]
	\begin{center}
		\includegraphics[width=0.5\textwidth]{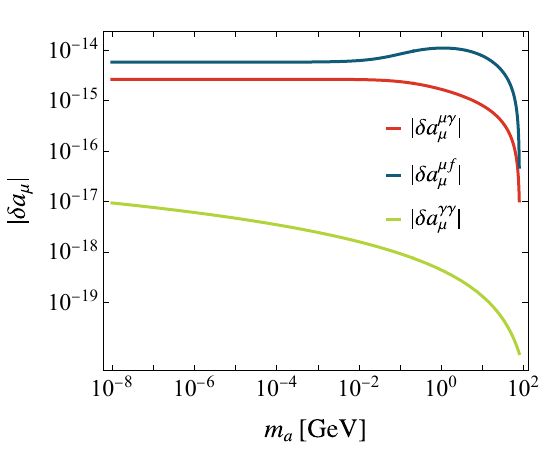}
		\caption{Comparison between the absolutes of different contributions in $\delta a_\mu$. Note that the LbL and photon polarization terms (proportional to $c_{\ga\ga}^2$) are several order of magnitude weaker than the other terms. We assumed $c/f=\SI{1}{\per\tera\electronvolt}$, where $c$ is any ALP-SM coupling.\label{fig:compgagaloop}}
	\end{center}
\end{figure}

We find it instructive to compare the effects from the different terms contributing to $\delta a_\mu$ (see Fig. \ref{fig:compamu} for comparison). For this reason we decompose $\delta a_\mu$ into
\begin{equation}
	\label{eq:decdela}
	\delta a_\mu=\frac{m_\mu^2}{f^2}K_{a\mu}(f,\mu)+\frac{1}{f^2}\left(c_{\mu\mu}^2\delta a_\mu^{\mu\mu}+c_{\mu\mu}\tilde c_{\ga\ga}\delta a_\mu^{\mu\ga}+c_{\mu\mu}c_{ff}\delta a_\mu^{\mu f}+c_{\ga\ga}^2\delta a_\mu^{\ga\ga}\right)\,.
\end{equation} 
Combinations of ALP-SM particle couplings missing here such as terms proportional to $c_{ff}^2$ are of higher loop order and thus further suppressed. We will neglect these terms here. The magnitude of the individual terms is compared in figure \ref{fig:compgagaloop}. Here we only keep leptons in the loop in $\delta a_\mu^{\mu f}$. The far dominant contribution over all ALP masses is $\delta a_\mu^{\mu\mu}$, exceeding the others by multiple orders of magnitude. For this reason we exclude this contribution in the overview for better visibility. It can be seen from this figure that the two-loop contribution is of roughly the same order as the one-loop contribution with a tree-level ALP-photon coupling and therefore should not be neglected when including the latter one. Note that the LbL and photon polarization diagrams combined (proportional to $c_{\ga\ga}^2$) are several orders of magnitude weaker than the $c_{\mu\mu}c_{ff}$ and $c_{\mu\mu}c_{\ga\ga}$ contribution.
The results worked out in this section can easily be transferred to the similar cases of the anomalous magnetic moment of the electron $a_e$ and, in principle, to that of the tau-lepton $a_\tau$. The latter is however experimentally too weakly constrained to give meaningful constraints on new physics models. Since it is not entirely clear what the Standard Model predicts for the value of $a_\mu$, we refrain from deriving bounds on the ALP couplings here.

\subsection[Application to $\mu\to e\ga$]{Application to \boldmath$\mu\to e\ga$}
An ALP with tree-level flavor-changing couplings can give rise to LFV decays like $\mu\to e\ga$. Unlike quark flavor-changing couplings, lepton flavor-changing couplings cannot be generated through evolution and matching corrections, thus they need to be imprinted already at the high energy scale, where the PQ symmetry is still unbroken \cite{Bauer:2020jbp,Chala:2020wvs}. Explicit models featuring such a flavor structure include for example axiflavon models, where the PQ symmetry is identified with the $U(1)$ symmetry of the Froggatt-Nielsen mechanism that could potentially explain the mass hierarchy of the SM fermions.

Diagrams contributing to the decay rate of $\mu\to e \ga$ are given in figure \ref{fig:2loop} for the case that an effective ALP-photon coupling is present. Additionally penguin-like diagrams where the ALP-loop closes on the fermion line also contribute. The branching ratio is given by
\begin{equation}
	\label{eq:brmeg}
	\text{Br}(\mu\to e\ga)=\frac{m_\mu^3}{8\pi\Gamma_\mu}\left(|F_2(0)|^2+|F_2^{5}(0)|^2\right)\,,
\end{equation}
and the form factors read
\begin{equation}
	\label{eq:ffsmeg}
	\begin{aligned}
		F_2^{(5)}(0)=&\frac{-e m_\mu}{64\pi}\frac{|k_E|_{e\mu }\mp|k_e|_{e\mu}}{f}\!\bigg(\!\!\frac{c_{\mu\mu}}{f}g_1\!\left(\frac{m_a^2}{m_\mu^2}\right)+\frac{\alpha}{\pi}\!\underbrace{\bigg[\frac{\tilde{c}_{\ga\ga}}{f}\mathcal{I}_1^{m_e\ll m_\mu}\!+\!\!\sum\limits_f\! N_c^f Q_f^2\frac{c_{ff}}{f}\mathcal{I}_2^{m_e\ll m_\mu}\bigg]}_{=c_{\ga\ga}^\text{eff}}\!\!\bigg),
	\end{aligned}
\end{equation}
where $\mathcal{I}_1^{m_e\ll m_\mu}$ and $\mathcal{I}_2^{m_e\ll m_\mu}$ have been given in eqs. \eqref{eq:ghdef} and \eqref{eq:I2limits}, respectively, and 
\begin{equation}
	\label{eq:g1def}
	g_1(x)=2x^{3/2}\sqrt{4-x}\arccos\frac{\sqrt{x}}{2}+1-2x+\frac{x^2(3-x)}{1-x}\ln x
\end{equation}
Note that by taking the limit $m_e\ll m_\mu$ we neglect terms that scale as $m_e^2/m_\mu^2=\mathcal{O}(\SI{e-5}{})$ in the branching ratio, making this a reasonable approximation.

\begin{figure}[t]
	\begin{center}
\begin{subfigure}[t]{0.45\textwidth}
	\centering
	\includegraphics[width=\textwidth]{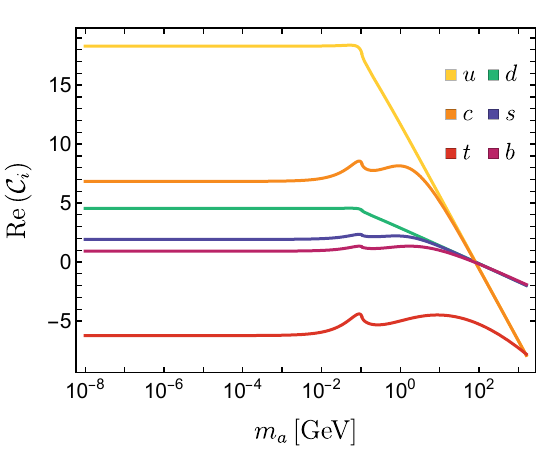}
\end{subfigure}
\begin{subfigure}[t]{0.45\textwidth}
	\centering
	\includegraphics[width=\textwidth]{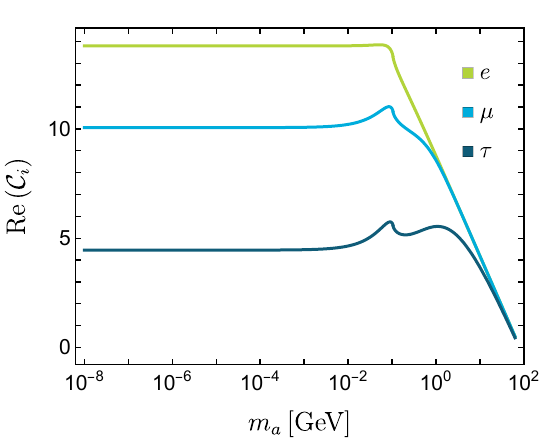}
\end{subfigure}
\caption{Effective ALP-photon coupling $c_{\ga\ga}^\text{eff}$ in $\mu\to e\ga$ for (left) a quark and (right) a lepton in the loop. For this plot we assumed $c_{ff}/f=\SI{1}{\per\tera\electronvolt}$. Note that for large ALP masses the functions become degenerate and the slope depends on the only on the number of colors and the electric charge of the fermions in the loop. \label{fig:effgaga}}
	\end{center}
\end{figure}

As before, we like to show the numerical implications of including the fermion-loop induced ALP-photon coupling to the effective ALP-photon coupling. We decompose the coupling into 
\begin{equation}
	\label{eq:cgagaeffdecompose2}
	c_{\ga\ga}^\text{eff}=c_{\ga\ga}\mathcal{I}_1^{m_2\ll m_1}+\sum\limits_i \mathcal{C}_i c_{ii}\,,
\end{equation}
in accordance with \eqref{eq:cgagaeffdecompose}. The results for the real parts of the coefficients are shown in figure \ref{fig:effgaga}. Overall, the picture looks very similar to the flavor-conserving case in figure \ref{fig:fermionloops}. While the effective coupling tends to a constant for ALPs that are very light, it yields a logarithmic function dependent only on the ALP mass with the slope in the logarithmic plot dictated by the number of colors of the inner loop fermion in the case that the ALP is heavy. Again, it is notable that light fermions in the loop give a larger contribution than heavy fermions. The cusps and wiggles of the curves at values $m_a\sim m_\mu$ are due to the fact that $c_{\ga\ga}^\text{eff}$ has imaginary values for ALPs lighter than the muon mass. We want to give a feeling of the size of the two-loop effects also in the flavor-changing case. If all $c_{ii}$ are set to zero and only $c_{\ga\ga}$ is kept, the real part of the effective ALP-photon coupling in the vanishing ALP mass limit reads $\lim\limits_{m_a\to 0}\text{Re}\,\,\mathcal{I}^{m_2\ll m_1}=13.8$. Comparing this value to the value of the plateaus for small ALP masses shows that also in ALP mediated flavor-changing form factors a loop-induced ALP-photon coupling is of the same order of magnitude than a tree-level one.

\begin{figure}[t]
	\begin{center}
		\includegraphics[width=0.495\textwidth]{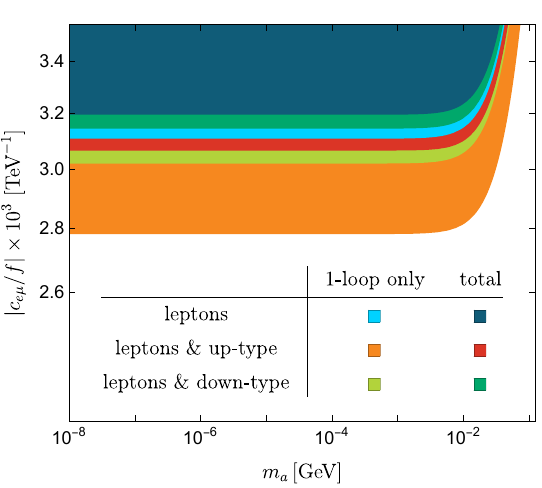}
		\includegraphics[width=0.495\textwidth]{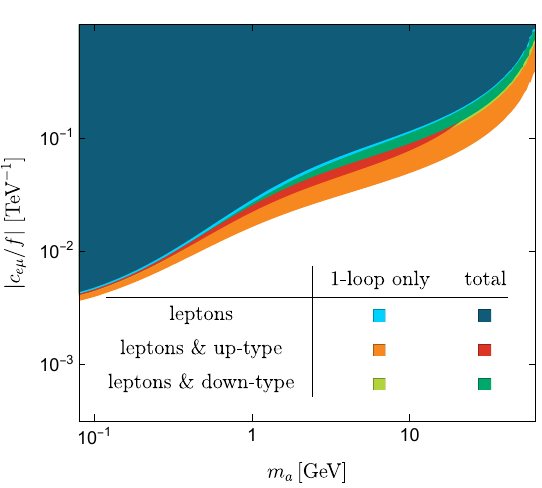}
		\caption{Parameter space excluded by the non-observation of $\mu\to e\ga$ for different coupling structures of the ALP, i.e. a coupling only to leptons (light blue and dark blue), leptons and up-type quarks (orange and red) and leptons and down-type quarks (light green and dark green). To visualize the effect of our findings, we show the limits for the given coupling structures on the flavor-violating ALP coupling $c_{e\mu}=\sqrt{|k_E|^2_{e\mu}+|k_e|^2_{e\mu}}$ when the two-loop contribution is neglected (first color) and taken into account (second color). The diagonal ALP couplings are assumed to be $c_{ff}/f=\SI{1}{\per\tera\eV}$. We show the constraints in the two regions $m_a\lesssim m_\mu$ (left) and $m_a\gtrsim m_\mu$ (right). For $m_a<m_\mu$ stronger bounds on $|c_{e\mu}|$ than shown here arise from experiments like $\mu\to3e$ and direct searches for $\mu\to ea$ decays. \label{fig:Brmegs}}
	\end{center}
\end{figure}

As an illustration how different coupling structures of the ALP imprint themselves on constraints on the ALP parameter space with one lepton flavor-changing coupling, we show the area of $c_{e\mu}$-- $m_a$ parameter space derived from the latest limit on the muon branching ratio $\text{Br}(\mu\to e\gamma)<\SI{4.2e-13}{}$ in figure \ref{fig:Brmegs}. The bound was obtained from the MEG collaboration \cite{TheMEG:2016wtm}. Here, the LFV ALP coupling is defined as $c_{e\mu}=\sqrt{|k_E|_{e\mu}^2+|k_e|_{e\mu}^2}$. We assume that additionally to the LFV coupling a diagonal coupling to the other leptons is present, too. In the plot we distinguish the following cases: In blue (yellow) we show the scenario where the ALP only couples to leptons, neglecting (taking into account) the additional two-loop piece from inserting a lepton loop in the effective ALP-photon coupling. The areas shaded in orange (red) and light green (dark green) represent the corresponding scenarios when the ALP features, additionally to the lepton coupling, a coupling to up-type quarks and down-type quarks, respectively. We keep the ALP mass and its LFV coupling as free parameters and set all lepton couplings universally to $c_{\ell\ell}/f=\SI{1}{\per\tera\eV}$. In a similar way, $c_{uu}/f=c_{cc}/f=c_{tt}/f=\SI{1}{\per\tera\eV}$ and $c_{dd}=c_{ss}=c_{bb}=\SI{1}{\per\tera\eV}$ are fixed in the second and third scenario, respectively. Note that taking the two-loop contributions into account generally leads to a weakening of the derived bounds. Even though the absolute bound on the LFV coupling does not change too much when the two-loop diagrams are considered because the dominant contribution is still given by the penguin-like diagrams proportional to $c_{e\mu}c_{\mu\mu}$, relative corrections of $\approx\SI{10}{\percent}$ are encountered.

Note that if the ALP mass is below the muon threshold, i.e.\ $m_a<m_\mu-m_e$, the search for $\mu\to e\gamma$ does not give rise to the strongest constraints on LFV ALP couplings. Stronger bounds arise from the indirect probes $\mu\to e a\to e\ga\ga$ and $\mu\to ea\to 3e$, where the ALP can be produced on-shell and subsequently decays into a photon (lepton) pair, as well as direct probes from searches for $\mu\to ea(\gamma)$. Limits derived from these experiments are typically many orders of magnitude stronger \cite{Bauer:2019gfk,Cornella:2019uxs,Bauer:2021mvw,Calibbi:2020jvd,Jho:2022snj,Knapen:2023zgi}. However, LFV couplings of heavy ALPs $m_a>m_\mu-m_e$ are best probed by $\mu\to e\gamma$.

\section{Generalization to gluons and application to the chromomagnetic moment of the top quark}
\label{sec:chromotop}
The computations presented for the anomalous magnetic moment of the muon in section \ref{sec:amm} are easily generalizable to non-abelian gauge bosons, giving rise to chromomagnetic moments of quarks. Especially the chromomagnetic dipole moment of the top quark $\hat{\mu}_t$ is an excellent probe for new physics above the electroweak scale \cite{Atwood:1994vm,Haberl:1995ek,Cheung:1995nt}. It is defined as the coefficient of the operator
\begin{equation}
	\label{eq:chromotopop}
	\mathcal{L}\supset -\hat{\mu}_t\frac{g_s}{2m_t}\bar{t}\sigma^{\mu\nu}T^a\,t\,G^a_{\mu\nu}\,.
\end{equation}
The latest result on its value was derived by the CMS collaboration in two independent measurements \cite{CMS:2019kzp,CMS:2019nrx} and yields the exclusion limit
\begin{equation}
	\label{eq:exmut}
	-0.014<\text{Re}(\hat{\mu}_t)<0.004
\end{equation}
at $\SI{95}{\percent}$ confidence level. The ALP could possibly induce a non-vanishing chromomagnetic moment via the diagrams given in figure \ref{fig:compamu}, where one has to exchange the photons by gluons. Its contribution is given by
\begin{equation}
	\label{eq:mut}
	\hat{\mu}_t=\frac{m_t^2}{32\pi^2f^2}\bigg\{c_{tt}^2h_1\left(\frac{m_a^2}{m_t^2}\right)+\frac{2\alpha_s}{\pi}c_{tt}\bigg[\tilde c_{GG}\mathcal{I}_1^{m_1=m_2}+\frac12\sum\limits_qc_{qq}\mathcal{I}_2^{m_1=m_2}\bigg]\bigg\}\,.
\end{equation}
The loop functions are given in equations \eqref{eq:h1def}, \eqref{eq:I1limits} and \eqref{eq:I2limits}, respectively. For ALPs that are significantly lighter than the top quark the two-loop contribution is given by \eqref{eq:limitallesschwer}. Terms proportional to $\tilde c_{GG}^2$ are suppressed by an additional factor of $\alpha_s^2$. Using RG methods, the leading logarithmic contributions have been first derived in \cite{Galda:2021hbr}. When we want to derive limits on the ALP parameters entering \eqref{eq:mut} from the experimental bound \eqref{eq:exmut} we must however mention an important caveat. The measurement puts constraints on the value of $\hat{\mu}_t$ through the operator \eqref{eq:chromotopop} solely. Any bound derived here therefore implicitly implies that the ALP's influence is mostly covered by aforementioned operator. This means especially that the ALP does not alter the underlying $pp\to t\bar{t}$ production process significantly. We consider this assumption justified for the heaviest ALPs $m_a\gtrsim m_\Upsilon\approx\SI{10.58}{\giga\eV}$. 

\begin{figure}[t]
	\begin{center}
		\includegraphics[width=0.6\textwidth]{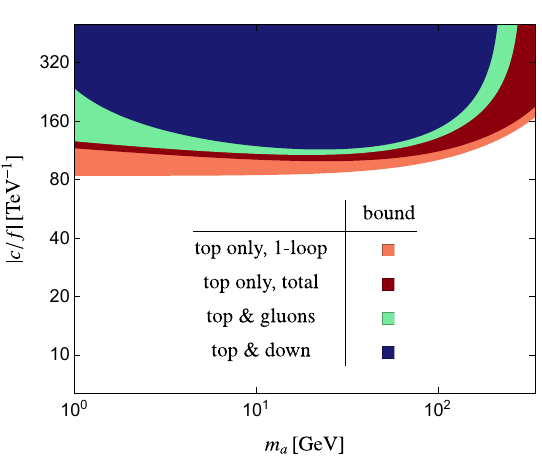}
		\caption{Parameter space excluded by searches for the chromomagnetic dipole moment of the top quark. Excluded regions are shown for the case when the ALP couples only to top quarks (light and dark red), additionally to gluons (green), and top quarks and down-type quarks (blue). All analysis assume that the couplings are of the same order $c_{tt}\sim c_{GG}\sim c_{qq}\sim c$. The scenario where the ALP couples uniformly to all up-type quarks coincides with the one where a coupling to top quarks and gluons is present, assuming all couplings are of the same order of magnitude. In the light red region we neglected the contribution from the two-loop graph, giving rise to an overestimation of the excluded parameter space of $\sim\SI{10}{\percent}$ for $m_a=\SI{100}{\giga\eV}$. \label{fig:chromotop}}
	\end{center}
\end{figure}

We show our derived bounds in figure \ref{fig:chromotop} for varying ALP masses for the three cases that the ALP couples only to the top quark (red), the top quark and gluons (green), and the top quark and down-type quarks (blue). In light red we show the derived bounds for top couplings only in the case that the two-loop contribution $\mathcal{I}_2^{m_1=m_2}$ is neglected. Note that similar to the $\mu\to e\gamma$ discussion including two-loop terms leads to a weakening of the derived constraint on the ALP parameter. We assume that all non-vanishing couplings are of the same order, i.e. $c_{tt}\sim c_{GG}\sim c_{qq}\sim c$. The case in which the ALP couples uniformly to all up-type quarks is not portrayed, because in the plot it would be indistinguishable from the second case (top and gluon coupling). This observation is easy to explain: The one-loop contribution is proportional to the coupling structure $\tilde c_{GG}=c_{GG}+1/2 \sum c_{qq}$. The coupling $c_{tt}$ must be present to generate a non-vanishing $\hat{\mu}_t$ in the first place. Hence, taking the additional couplings to up and charm quarks into account has the same effect as keeping a tree-level ALP-gluon coupling, for the assumption that all ALP couplings are of the same order. For all quarks other than the top quark the two-loop contribution $\mathcal{I}_2^{m_1=m_2}$ is negligible. Comparing the two red regions in the plot, we emphasize that the two-loop corrections entering the effective ALP-gluon coupling should not be neglected. For an ALP with a mass of $m_a=\SI{100}{\giga\eV}$, the corrections to the derived bound are of the order of $\sim\SI{10}{\percent}$.

\section{Conclusions}
Axion-like particles are among the best motivated models for physics beyond the Standard Model. Since they arise as pseudo-Nambu-Goldstone bosons of a high-energy spontaneously broken $U(1)$ symmetry, they are part of many models. Furthermore they have the potential to tackle multiple issues of the SM at once, such as the strong CP problem and the absence of dark matter in the SM. In an often used formulation of the most general ALP-Lagrangian the Wilson coefficient associated with the ALP-photon coupling (or a general ALP to gauge boson coupling) is suppressed by a factor of $\alpha/(4\pi)$. The advantage of this normalization is that up to two-loop order the gauge-boson couplings of the ALP are scale independent. However, this comes at the price that when the ALP-photon coupling is replaced by an ALP coupling to a fermion loop that radiates the two photons, this additional loop contributes with the same power of the QED coupling constant. For a consistent treatment of ALP-photon couplings it is therefore necessary to include this contribution into the computations of Feynman diagrams.

In this work we have studied the impact of the added fermion loop contribution to $\psi_1\to \psi_2\ga$ form factors. Examples where such form factors appear in phenomenological studies are the lepton flavor-violating decay $\mu\to e\ga$, as well as the flavor-conserving observable of the anomalous magnetic moment of the muon $a_\mu$. For the physically relevant cases that either the two fermions $\psi_1$ and $\psi_2$ are the same, or that the final state fermion is much lighter than the initial state one, we were able to derive explicit expressions for both the one-loop as well as the fermion-loop induced two-loop contribution to the form factor. The mass hierarchy between the different fermion families ensures that the assumption $m_2\ll m_1$ in the flavor-violating case is indeed justified. We find that generally for both flavor-conserving and -violating processes the two-loop piece shows the same behavior in certain mass limits: If the ALP is much heavier than the loop fermion, the latter decouples in the Barr-Zee graphs and in essence the virtual loop momenta are cut off for values below the highest mass scale in the process. If on the other hand the inner loop fermion is much heavier than both the ALP and the initial state fermion, the loop fermion can be integrated out.

Eventually, we applied our findings to the two lepton observables of $(g-2)_\mu$ and $\mu\to e \gamma$. We find for both scenarios that the effective ALP-photon coupling including the one and two-loop contributions shows a similar behavior. When the inner fermion is heavy, the ALP-photon coupling tends to a constant dependent only on the fermion's mass. When the ALP is the heaviest particle instead, the effective coupling is a linear function of the ALP mass in logarithmic plots and the slope is dictated by the number of flavors in the loop. We showed that neglecting the two-loop contribution results in an overestimation of the effects of up to $\approx\SI{10}{\percent}$. For the LFV observable $\mu\to e\gamma$ we presented the impact of this finding on the excluded parameter space in the three scenarios where the ALP features additionally to the LFV coupling a coupling to leptons, leptons and up-type quarks, and leptons and down-type quarks, respectively. A similar analysis is currently not possible for the anomalous magnetic moment of the muon, because it is unclear what the SM predicted value is due to the discrepancy of the two generally used methods to extract the hadronic vacuum polarization, i.e. the term that is currently driving the SM error budget. 

In the last section we eventually generalized our results to the study of the chromomagnetic moment of the top quark. We found that the insertion of the fermion loop has an impact only when the fermion in the loop itself is a top quark, too. Neglecting this effect however gives rise to an overestimation of the excluded parameter space of about $\SI{10}{\percent}$. 
	
\section*{Acknowledgments}
M.S. would like to thank Claudia Cornella for useful discussions. The research of M.N. and M.S. are supported by the Cluster of Excellence \textit{Precision Physics, Fundamental Interactions and Structure of Matter} (PRISMA$^+$ -- EXC 2118/1) within the German Excellence Strategy (project ID 390831469). M.S. gratefully acknowledges support from the Alexander von Humboldt Foundation as a Feodor Lynen Fellow. M.S. is supported by the United States Department of Energy under Grant Contract No. DESC0012704.
	
	\addcontentsline{toc}{section}{References}
	\bibliographystyle{JHEP}
	\bibliography{references}

\providecommand{\href}[2]{#2}\begingroup\raggedright\begin{thebibliography}{10}

\bibitem{Peccei:1977hh}
R.~D. Peccei and H.~R. Quinn, \emph{{CP Conservation in the Presence of
  Instantons}}, \href{https://doi.org/10.1103/PhysRevLett.38.1440}{\emph{Phys.
  Rev. Lett.} {\bfseries 38} (1977) 1440--1443}.

\bibitem{Peccei:1977ur}
R.~D. Peccei and H.~R. Quinn, \emph{{Constraints Imposed by CP Conservation in
  the Presence of Instantons}},
  \href{https://doi.org/10.1103/PhysRevD.16.1791}{\emph{Phys. Rev.} {\bfseries
  D16} (1977) 1791--1797}.

\bibitem{Weinberg:1977ma}
S.~Weinberg, \emph{{A New Light Boson?}},
  \href{https://doi.org/10.1103/PhysRevLett.40.223}{\emph{Phys. Rev. Lett.}
  {\bfseries 40} (1978) 223--226}.

\bibitem{Wilczek:1977pj}
F.~Wilczek, \emph{{Problem of Strong $P$ and $T$ Invariance in the Presence of
  Instantons}}, \href{https://doi.org/10.1103/PhysRevLett.40.279}{\emph{Phys.
  Rev. Lett.} {\bfseries 40} (1978) 279--282}.

\bibitem{Ema:2016ops}
Y.~Ema, K.~Hamaguchi, T.~Moroi and K.~Nakayama, \emph{{Flaxion: a minimal
  extension to solve puzzles in the standard model}},
  \href{https://doi.org/10.1007/JHEP01(2017)096}{\emph{JHEP} {\bfseries 01}
  (2017) 096}, [\href{https://arxiv.org/abs/1612.05492}{{\ttfamily
  1612.05492}}].

\bibitem{Calibbi:2016hwq}
L.~Calibbi, F.~Goertz, D.~Redigolo, R.~Ziegler and J.~Zupan, \emph{{Minimal
  axion model from flavor}},
  \href{https://doi.org/10.1103/PhysRevD.95.095009}{\emph{Phys. Rev.}
  {\bfseries D95} (2017) 095009},
  [\href{https://arxiv.org/abs/1612.08040}{{\ttfamily 1612.08040}}].

\bibitem{Alanne:2018fns}
T.~Alanne, S.~Blasi and F.~Goertz, \emph{{Common source for scalars: Flavored
  axion-Higgs unification}},
  \href{https://doi.org/10.1103/PhysRevD.99.015028}{\emph{Phys. Rev. D}
  {\bfseries 99} (2019) 015028},
  [\href{https://arxiv.org/abs/1807.10156}{{\ttfamily 1807.10156}}].

\bibitem{Ringwald:2016yge}
A.~Ringwald, \emph{{Alternative dark matter candidates: Axions}},
  \href{https://doi.org/10.22323/1.283.0081}{\emph{PoS} {\bfseries NOW2016}
  (2016) 081}, [\href{https://arxiv.org/abs/1612.08933}{{\ttfamily
  1612.08933}}].

\bibitem{Machado:2018nqk}
C.~S. Machado, W.~Ratzinger, P.~Schwaller and B.~A. Stefanek, \emph{{Audible
  Axions}}, \href{https://doi.org/10.1007/JHEP01(2019)053}{\emph{JHEP}
  {\bfseries 01} (2019) 053},
  [\href{https://arxiv.org/abs/1811.01950}{{\ttfamily 1811.01950}}].

\bibitem{Machado:2019xuc}
C.~S. Machado, W.~Ratzinger, P.~Schwaller and B.~A. Stefanek,
  \emph{{Gravitational wave probes of axionlike particles}},
  \href{https://doi.org/10.1103/PhysRevD.102.075033}{\emph{Phys. Rev. D}
  {\bfseries 102} (2020) 075033},
  [\href{https://arxiv.org/abs/1912.01007}{{\ttfamily 1912.01007}}].

\bibitem{Madge:2021abk}
E.~Madge, W.~Ratzinger, D.~Schmitt and P.~Schwaller, \emph{{Audible axions with
  a booster: Stochastic gravitational waves from rotating ALPs}},
  \href{https://doi.org/10.21468/SciPostPhys.12.5.171}{\emph{SciPost Phys.}
  {\bfseries 12} (2022) 171},
  [\href{https://arxiv.org/abs/2111.12730}{{\ttfamily 2111.12730}}].

\bibitem{Kim:1979if}
J.~E. Kim, \emph{{Weak Interaction Singlet and Strong CP Invariance}},
  \href{https://doi.org/10.1103/PhysRevLett.43.103}{\emph{Phys. Rev. Lett.}
  {\bfseries 43} (1979) 103}.

\bibitem{Shifman:1979if}
M.~A. Shifman, A.~I. Vainshtein and V.~I. Zakharov, \emph{{Can Confinement
  Ensure Natural CP Invariance of Strong Interactions?}},
  \href{https://doi.org/10.1016/0550-3213(80)90209-6}{\emph{Nucl. Phys. B}
  {\bfseries 166} (1980) 493--506}.

\bibitem{Dine:1981rt}
M.~Dine, W.~Fischler and M.~Srednicki, \emph{{A Simple Solution to the Strong
  CP Problem with a Harmless Axion}},
  \href{https://doi.org/10.1016/0370-2693(81)90590-6}{\emph{Phys. Lett. B}
  {\bfseries 104} (1981) 199--202}.

\bibitem{Zhitnitsky:1980tq}
A.~R. Zhitnitsky, \emph{{On Possible Suppression of the Axion Hadron
  Interactions. (In Russian)}}, {\emph{Sov. J. Nucl. Phys.} {\bfseries 31}
  (1980) 260}.

\bibitem{Ringwald:2012hr}
A.~Ringwald, \emph{{Exploring the Role of Axions and Other WISPs in the Dark
  Universe}}, \href{https://doi.org/10.1016/j.dark.2012.10.008}{\emph{Phys.
  Dark Univ.} {\bfseries 1} (2012) 116--135},
  [\href{https://arxiv.org/abs/1210.5081}{{\ttfamily 1210.5081}}].

\bibitem{Bauer:2020jbp}
M.~Bauer, M.~Neubert, S.~Renner, M.~Schnubel and A.~Thamm, \emph{{The
  Low-Energy Effective Theory of Axions and ALPs}},
  \href{https://doi.org/10.1007/JHEP04(2021)063}{\emph{JHEP} {\bfseries 04}
  (2021) 063}, [\href{https://arxiv.org/abs/2012.12272}{{\ttfamily
  2012.12272}}].

\bibitem{Bauer:2021mvw}
M.~Bauer, M.~Neubert, S.~Renner, M.~Schnubel and A.~Thamm, \emph{{Flavor probes
  of axion-like particles}},
  \href{https://arxiv.org/abs/2110.10698}{{\ttfamily 2110.10698}}.

\bibitem{Chala:2020wvs}
M.~Chala, G.~Guedes, M.~Ramos and J.~Santiago, \emph{{Running in the ALPs}},
  \href{https://doi.org/10.1140/epjc/s10052-021-08968-2}{\emph{Eur. Phys. J. C}
  {\bfseries 81} (2021) 181},
  [\href{https://arxiv.org/abs/2012.09017}{{\ttfamily 2012.09017}}].

\bibitem{Davidson:1981zd}
A.~Davidson and K.~C. Wali, \emph{{MINIMAL FLAVOR UNIFICATION VIA
  MULTIGENERATIONAL PECCEI-QUINN SYMMETRY}},
  \href{https://doi.org/10.1103/PhysRevLett.48.11}{\emph{Phys. Rev. Lett.}
  {\bfseries 48} (1982) 11}.

\bibitem{Davidson:1984ik}
A.~Davidson and M.~A.~H. Vozmediano, \emph{{The Horizontal Axion Alternative:
  The Interplay of Vacuum Structure and Flavor Interactions}},
  \href{https://doi.org/10.1016/0550-3213(84)90616-3}{\emph{Nucl. Phys. B}
  {\bfseries 248} (1984) 647--670}.

\bibitem{Peccei:1986pn}
R.~D. Peccei, T.~T. Wu and T.~Yanagida, \emph{{A VIABLE AXION MODEL}},
  \href{https://doi.org/10.1016/0370-2693(86)90284-4}{\emph{Phys. Lett. B}
  {\bfseries 172} (1986) 435--440}.

\bibitem{Krauss:1986wx}
L.~M. Krauss and F.~Wilczek, \emph{{A SHORTLIVED AXION VARIANT}},
  \href{https://doi.org/10.1016/0370-2693(86)90244-3}{\emph{Phys. Lett. B}
  {\bfseries 173} (1986) 189--192}.

\bibitem{Geng:1988nc}
C.~Q. Geng and J.~N. Ng, \emph{{Flavor Connections and Neutrino Mass Hierarchy
  Invariant Invisible Axion Models Without Domain Wall Problem}},
  \href{https://doi.org/10.1103/PhysRevD.39.1449}{\emph{Phys. Rev. D}
  {\bfseries 39} (1989) 1449}.

\bibitem{Celis:2014iua}
A.~Celis, J.~Fuentes-Martin and H.~Serodio, \emph{{An invisible axion model
  with controlled FCNCs at tree level}},
  \href{https://doi.org/10.1016/j.physletb.2014.12.028}{\emph{Phys. Lett. B}
  {\bfseries 741} (2015) 117--123},
  [\href{https://arxiv.org/abs/1410.6217}{{\ttfamily 1410.6217}}].

\bibitem{Alves:2017avw}
D.~S.~M. Alves and N.~Weiner, \emph{{A viable QCD axion in the MeV mass
  range}}, \href{https://doi.org/10.1007/JHEP07(2018)092}{\emph{JHEP}
  {\bfseries 07} (2018) 092},
  [\href{https://arxiv.org/abs/1710.03764}{{\ttfamily 1710.03764}}].

\bibitem{DiLuzio:2017ogq}
L.~Di~Luzio, F.~Mescia, E.~Nardi, P.~Panci and R.~Ziegler, \emph{{Astrophobic
  Axions}}, \href{https://doi.org/10.1103/PhysRevLett.120.261803}{\emph{Phys.
  Rev. Lett.} {\bfseries 120} (2018) 261803},
  [\href{https://arxiv.org/abs/1712.04940}{{\ttfamily 1712.04940}}].

\bibitem{Choi:2017gpf}
K.~Choi, S.~H. Im, C.~B. Park and S.~Yun, \emph{{Minimal Flavor Violation with
  Axion-like Particles}},
  \href{https://doi.org/10.1007/JHEP11(2017)070}{\emph{JHEP} {\bfseries 11}
  (2017) 070}, [\href{https://arxiv.org/abs/1708.00021}{{\ttfamily
  1708.00021}}].

\bibitem{MartinCamalich:2020dfe}
J.~Martin~Camalich, M.~Pospelov, P.~N.~H. Vuong, R.~Ziegler and J.~Zupan,
  \emph{{Quark Flavor Phenomenology of the QCD Axion}},
  \href{https://doi.org/10.1103/PhysRevD.102.015023}{\emph{Phys. Rev. D}
  {\bfseries 102} (2020) 015023},
  [\href{https://arxiv.org/abs/2002.04623}{{\ttfamily 2002.04623}}].

\bibitem{Gelmini:1982zz}
G.~B. Gelmini, S.~Nussinov and T.~Yanagida, \emph{{Does Nature Like
  Nambu-Goldstone Bosons?}},
  \href{https://doi.org/10.1016/0550-3213(83)90426-1}{\emph{Nucl. Phys. B}
  {\bfseries 219} (1983) 31--40}.

\bibitem{Anselm:1985bp}
A.~A. Anselm, N.~G. Uraltsev and M.~Y. Khlopov, \emph{{mu ---\ensuremath{>} e
  FAMILON DECAY}}, {\emph{Sov. J. Nucl. Phys.} {\bfseries 41} (1985) 1060}.

\bibitem{Bauer:2017nlg}
M.~Bauer, M.~Neubert and A.~Thamm, \emph{{LHC as an Axion Factory: Probing an
  Axion Explanation for $(g-2)_\mu$ with Exotic Higgs Decays}},
  \href{https://doi.org/10.1103/PhysRevLett.119.031802}{\emph{Phys. Rev. Lett.}
  {\bfseries 119} (2017) 031802},
  [\href{https://arxiv.org/abs/1704.08207}{{\ttfamily 1704.08207}}].

\bibitem{Bauer:2017ris}
M.~Bauer, M.~Neubert and A.~Thamm, \emph{{Collider Probes of Axion-Like
  Particles}}, \href{https://doi.org/10.1007/JHEP12(2017)044}{\emph{JHEP}
  {\bfseries 12} (2017) 044},
  [\href{https://arxiv.org/abs/1708.00443}{{\ttfamily 1708.00443}}].

\bibitem{Bauer:2019gfk}
M.~Bauer, M.~Neubert, S.~Renner, M.~Schnubel and A.~Thamm, \emph{{Axionlike
  Particles, Lepton-Flavor Violation, and a New Explanation of $a_\mu$ and
  $a_e$}}, \href{https://doi.org/10.1103/PhysRevLett.124.211803}{\emph{Phys.
  Rev. Lett.} {\bfseries 124} (2020) 211803},
  [\href{https://arxiv.org/abs/1908.00008}{{\ttfamily 1908.00008}}].

\bibitem{Bauer:2021wjo}
M.~Bauer, M.~Neubert, S.~Renner, M.~Schnubel and A.~Thamm, \emph{{Consistent
  Treatment of Axions in the Weak Chiral Lagrangian}},
  \href{https://doi.org/10.1103/PhysRevLett.127.081803}{\emph{Phys. Rev. Lett.}
  {\bfseries 127} (2021) 081803},
  [\href{https://arxiv.org/abs/2102.13112}{{\ttfamily 2102.13112}}].

\bibitem{Buen-Abad:2021fwq}
M.~A. Buen-Abad, J.~Fan, M.~Reece and C.~Sun, \emph{{Challenges for an axion
  explanation of the muon $g - 2$ measurement}},
  \href{https://doi.org/10.1007/JHEP09(2021)101}{\emph{JHEP} {\bfseries 09}
  (2021) 101}, [\href{https://arxiv.org/abs/2104.03267}{{\ttfamily
  2104.03267}}].

\bibitem{Bardeen:1978nq}
W.~A. Bardeen, S.~H.~H. Tye and J.~A.~M. Vermaseren, \emph{{Phenomenology of
  the New Light Higgs Boson Search}},
  \href{https://doi.org/10.1016/0370-2693(78)90859-6}{\emph{Phys. Lett. B}
  {\bfseries 76} (1978) 580--584}.

\bibitem{DiVecchia:1980yfw}
P.~Di~Vecchia and G.~Veneziano, \emph{{Chiral Dynamics in the Large n Limit}},
  \href{https://doi.org/10.1016/0550-3213(80)90370-3}{\emph{Nucl. Phys. B}
  {\bfseries 171} (1980) 253--272}.

\bibitem{Cornella:2019uxs}
C.~Cornella, P.~Paradisi and O.~Sumensari, \emph{{Hunting for ALPs with Lepton
  Flavor Violation}},
  \href{https://doi.org/10.1007/JHEP01(2020)158}{\emph{JHEP} {\bfseries 01}
  (2020) 158}, [\href{https://arxiv.org/abs/1911.06279}{{\ttfamily
  1911.06279}}].

\bibitem{Chetyrkin:1998mw}
K.~G. Chetyrkin, B.~A. Kniehl, M.~Steinhauser and W.~A. Bardeen,
  \emph{{Effective QCD interactions of CP odd Higgs bosons at three loops}},
  \href{https://doi.org/10.1016/S0550-3213(98)00594-X}{\emph{Nucl. Phys. B}
  {\bfseries 535} (1998) 3--18},
  [\href{https://arxiv.org/abs/hep-ph/9807241}{{\ttfamily hep-ph/9807241}}].

\bibitem{Barr:1990vd}
S.~M. Barr and A.~Zee, \emph{{Electric Dipole Moment of the Electron and of the
  Neutron}}, \href{https://doi.org/10.1103/PhysRevLett.65.21}{\emph{Phys. Rev.
  Lett.} {\bfseries 65} (1990) 21--24}.

\bibitem{Galda:2023qjx}
A.~M. Galda and M.~Neubert, \emph{{ALP-LEFT Interference and the Muon
  $(g-2)$}}, \href{https://doi.org/10.1007/JHEP11(2023)015}{\emph{JHEP}
  {\bfseries 11} (2023) 015},
  [\href{https://arxiv.org/abs/2308.01338}{{\ttfamily 2308.01338}}].

\bibitem{Marciano:2016yhf}
W.~J. Marciano, A.~Masiero, P.~Paradisi and M.~Passera, \emph{{Contributions of
  axionlike particles to lepton dipole moments}},
  \href{https://doi.org/10.1103/PhysRevD.94.115033}{\emph{Phys. Rev. D}
  {\bfseries 94} (2016) 115033},
  [\href{https://arxiv.org/abs/1607.01022}{{\ttfamily 1607.01022}}].

\bibitem{Muong-2:2006rrc}
{\scshape Muon g-2} collaboration, G.~W. Bennett et~al., \emph{{Final Report of
  the Muon E821 Anomalous Magnetic Moment Measurement at BNL}},
  \href{https://doi.org/10.1103/PhysRevD.73.072003}{\emph{Phys. Rev. D}
  {\bfseries 73} (2006) 072003},
  [\href{https://arxiv.org/abs/hep-ex/0602035}{{\ttfamily hep-ex/0602035}}].

\bibitem{Muong-2:2021ojo}
{\scshape Muon g-2} collaboration, B.~Abi et~al., \emph{{Measurement of the
  Positive Muon Anomalous Magnetic Moment to 0.46 ppm}},
  \href{https://doi.org/10.1103/PhysRevLett.126.141801}{\emph{Phys. Rev. Lett.}
  {\bfseries 126} (2021) 141801},
  [\href{https://arxiv.org/abs/2104.03281}{{\ttfamily 2104.03281}}].

\bibitem{Aoyama:2020ynm}
T.~Aoyama et~al., \emph{{The anomalous magnetic moment of the muon in the
  Standard Model}},
  \href{https://doi.org/10.1016/j.physrep.2020.07.006}{\emph{Phys. Rept.}
  {\bfseries 887} (2020) 1--166},
  [\href{https://arxiv.org/abs/2006.04822}{{\ttfamily 2006.04822}}].

\bibitem{Davier:2010nc}
M.~Davier, A.~Hoecker, B.~Malaescu and Z.~Zhang, \emph{{Reevaluation of the
  Hadronic Contributions to the Muon g-2 and to alpha(MZ)}},
  \href{https://doi.org/10.1140/epjc/s10052-012-1874-8}{\emph{Eur. Phys. J. C}
  {\bfseries 71} (2011) 1515},
  [\href{https://arxiv.org/abs/1010.4180}{{\ttfamily 1010.4180}}].

\bibitem{Aoyama:2012wk}
T.~Aoyama, M.~Hayakawa, T.~Kinoshita and M.~Nio, \emph{{Complete Tenth-Order
  QED Contribution to the Muon g-2}},
  \href{https://doi.org/10.1103/PhysRevLett.109.111808}{\emph{Phys. Rev. Lett.}
  {\bfseries 109} (2012) 111808},
  [\href{https://arxiv.org/abs/1205.5370}{{\ttfamily 1205.5370}}].

\bibitem{Aoyama:2019ryr}
T.~Aoyama, T.~Kinoshita and M.~Nio, \emph{{Theory of the Anomalous Magnetic
  Moment of the Electron}},
  \href{https://doi.org/10.3390/atoms7010028}{\emph{Atoms} {\bfseries 7} (2019)
  28}.

\bibitem{Czarnecki:2002nt}
A.~Czarnecki, W.~J. Marciano and A.~Vainshtein, \emph{{Refinements in
  electroweak contributions to the muon anomalous magnetic moment}},
  \href{https://doi.org/10.1103/PhysRevD.67.073006}{\emph{Phys. Rev. D}
  {\bfseries 67} (2003) 073006},
  [\href{https://arxiv.org/abs/hep-ph/0212229}{{\ttfamily hep-ph/0212229}}].

\bibitem{Gnendiger:2013pva}
C.~Gnendiger, D.~St\"ockinger and H.~St\"ockinger-Kim, \emph{{The electroweak
  contributions to $(g-2)_\mu$ after the Higgs boson mass measurement}},
  \href{https://doi.org/10.1103/PhysRevD.88.053005}{\emph{Phys. Rev. D}
  {\bfseries 88} (2013) 053005},
  [\href{https://arxiv.org/abs/1306.5546}{{\ttfamily 1306.5546}}].

\bibitem{Davier:2017zfy}
M.~Davier, A.~Hoecker, B.~Malaescu and Z.~Zhang, \emph{{Reevaluation of the
  hadronic vacuum polarisation contributions to the Standard Model predictions
  of the muon $g-2$ and ${\alpha (m_Z^2)}$ using newest hadronic cross-section
  data}}, \href{https://doi.org/10.1140/epjc/s10052-017-5161-6}{\emph{Eur.
  Phys. J. C} {\bfseries 77} (2017) 827},
  [\href{https://arxiv.org/abs/1706.09436}{{\ttfamily 1706.09436}}].

\bibitem{Keshavarzi:2018mgv}
A.~Keshavarzi, D.~Nomura and T.~Teubner, \emph{{Muon $g-2$ and $\alpha(M_Z^2)$:
  a new data-based analysis}},
  \href{https://doi.org/10.1103/PhysRevD.97.114025}{\emph{Phys. Rev. D}
  {\bfseries 97} (2018) 114025},
  [\href{https://arxiv.org/abs/1802.02995}{{\ttfamily 1802.02995}}].

\bibitem{Colangelo:2018mtw}
G.~Colangelo, M.~Hoferichter and P.~Stoffer, \emph{{Two-pion contribution to
  hadronic vacuum polarization}},
  \href{https://doi.org/10.1007/JHEP02(2019)006}{\emph{JHEP} {\bfseries 02}
  (2019) 006}, [\href{https://arxiv.org/abs/1810.00007}{{\ttfamily
  1810.00007}}].

\bibitem{Hoferichter:2019gzf}
M.~Hoferichter, B.-L. Hoid and B.~Kubis, \emph{{Three-pion contribution to
  hadronic vacuum polarization}},
  \href{https://doi.org/10.1007/JHEP08(2019)137}{\emph{JHEP} {\bfseries 08}
  (2019) 137}, [\href{https://arxiv.org/abs/1907.01556}{{\ttfamily
  1907.01556}}].

\bibitem{Davier:2019can}
M.~Davier, A.~Hoecker, B.~Malaescu and Z.~Zhang, \emph{{A new evaluation of the
  hadronic vacuum polarisation contributions to the muon anomalous magnetic
  moment and to $\mathbf{\boldsymbol\alpha(m_Z^2)}$}},
  \href{https://doi.org/10.1140/epjc/s10052-020-7792-2}{\emph{Eur. Phys. J. C}
  {\bfseries 80} (2020) 241},
  [\href{https://arxiv.org/abs/1908.00921}{{\ttfamily 1908.00921}}].

\bibitem{Keshavarzi:2019abf}
A.~Keshavarzi, D.~Nomura and T.~Teubner, \emph{{$g-2$ of charged leptons,
  $\alpha (M^2_Z)$ , and the hyperfine splitting of muonium}},
  \href{https://doi.org/10.1103/PhysRevD.101.014029}{\emph{Phys. Rev. D}
  {\bfseries 101} (2020) 014029},
  [\href{https://arxiv.org/abs/1911.00367}{{\ttfamily 1911.00367}}].

\bibitem{Kurz:2014wya}
A.~Kurz, T.~Liu, P.~Marquard and M.~Steinhauser, \emph{{Hadronic contribution
  to the muon anomalous magnetic moment to next-to-next-to-leading order}},
  \href{https://doi.org/10.1016/j.physletb.2014.05.043}{\emph{Phys. Lett. B}
  {\bfseries 734} (2014) 144--147},
  [\href{https://arxiv.org/abs/1403.6400}{{\ttfamily 1403.6400}}].

\bibitem{Melnikov:2003xd}
K.~Melnikov and A.~Vainshtein, \emph{{Hadronic light-by-light scattering
  contribution to the muon anomalous magnetic moment revisited}},
  \href{https://doi.org/10.1103/PhysRevD.70.113006}{\emph{Phys. Rev. D}
  {\bfseries 70} (2004) 113006},
  [\href{https://arxiv.org/abs/hep-ph/0312226}{{\ttfamily hep-ph/0312226}}].

\bibitem{Masjuan:2017tvw}
P.~Masjuan and P.~Sanchez-Puertas, \emph{{Pseudoscalar-pole contribution to the
  $(g_{\mu}-2)$: a rational approach}},
  \href{https://doi.org/10.1103/PhysRevD.95.054026}{\emph{Phys. Rev. D}
  {\bfseries 95} (2017) 054026},
  [\href{https://arxiv.org/abs/1701.05829}{{\ttfamily 1701.05829}}].

\bibitem{Colangelo:2017fiz}
G.~Colangelo, M.~Hoferichter, M.~Procura and P.~Stoffer, \emph{{Dispersion
  relation for hadronic light-by-light scattering: two-pion contributions}},
  \href{https://doi.org/10.1007/JHEP04(2017)161}{\emph{JHEP} {\bfseries 04}
  (2017) 161}, [\href{https://arxiv.org/abs/1702.07347}{{\ttfamily
  1702.07347}}].

\bibitem{Hoferichter:2018kwz}
M.~Hoferichter, B.-L. Hoid, B.~Kubis, S.~Leupold and S.~P. Schneider,
  \emph{{Dispersion relation for hadronic light-by-light scattering: pion
  pole}}, \href{https://doi.org/10.1007/JHEP10(2018)141}{\emph{JHEP} {\bfseries
  10} (2018) 141}, [\href{https://arxiv.org/abs/1808.04823}{{\ttfamily
  1808.04823}}].

\bibitem{Gerardin:2019vio}
A.~G\'erardin, H.~B. Meyer and A.~Nyffeler, \emph{{Lattice calculation of the
  pion transition form factor with $N_f=2+1$ Wilson quarks}},
  \href{https://doi.org/10.1103/PhysRevD.100.034520}{\emph{Phys. Rev. D}
  {\bfseries 100} (2019) 034520},
  [\href{https://arxiv.org/abs/1903.09471}{{\ttfamily 1903.09471}}].

\bibitem{Bijnens:2019ghy}
J.~Bijnens, N.~Hermansson-Truedsson and A.~Rodr\'\i{}guez-S\'anchez,
  \emph{{Short-distance constraints for the HLbL contribution to the muon
  anomalous magnetic moment}},
  \href{https://doi.org/10.1016/j.physletb.2019.134994}{\emph{Phys. Lett. B}
  {\bfseries 798} (2019) 134994},
  [\href{https://arxiv.org/abs/1908.03331}{{\ttfamily 1908.03331}}].

\bibitem{Colangelo:2019uex}
G.~Colangelo, F.~Hagelstein, M.~Hoferichter, L.~Laub and P.~Stoffer,
  \emph{{Longitudinal short-distance constraints for the hadronic
  light-by-light contribution to $(g-2)_\mu$ with large-$N_c$ Regge models}},
  \href{https://doi.org/10.1007/JHEP03(2020)101}{\emph{JHEP} {\bfseries 03}
  (2020) 101}, [\href{https://arxiv.org/abs/1910.13432}{{\ttfamily
  1910.13432}}].

\bibitem{Blum:2019ugy}
T.~Blum, N.~Christ, M.~Hayakawa, T.~Izubuchi, L.~Jin, C.~Jung et~al.,
  \emph{{Hadronic Light-by-Light Scattering Contribution to the Muon Anomalous
  Magnetic Moment from Lattice QCD}},
  \href{https://doi.org/10.1103/PhysRevLett.124.132002}{\emph{Phys. Rev. Lett.}
  {\bfseries 124} (2020) 132002},
  [\href{https://arxiv.org/abs/1911.08123}{{\ttfamily 1911.08123}}].

\bibitem{Colangelo:2014qya}
G.~Colangelo, M.~Hoferichter, A.~Nyffeler, M.~Passera and P.~Stoffer,
  \emph{{Remarks on higher-order hadronic corrections to the muon
  g\ensuremath{-}2}},
  \href{https://doi.org/10.1016/j.physletb.2014.06.012}{\emph{Phys. Lett. B}
  {\bfseries 735} (2014) 90--91},
  [\href{https://arxiv.org/abs/1403.7512}{{\ttfamily 1403.7512}}].

\bibitem{Borsanyi:2020mff}
S.~Borsanyi et~al., \emph{{Leading hadronic contribution to the muon magnetic
  moment from lattice QCD}},
  \href{https://doi.org/10.1038/s41586-021-03418-1}{\emph{Nature} {\bfseries
  593} (2021) 51--55}, [\href{https://arxiv.org/abs/2002.12347}{{\ttfamily
  2002.12347}}].

\bibitem{TheMEG:2016wtm}
{\scshape MEG} collaboration, A.~M. Baldini et~al., \emph{{Search for the
  lepton flavour violating decay $\mu ^+ \rightarrow \mathrm {e}^+ \gamma $
  with the full dataset of the MEG experiment}},
  \href{https://doi.org/10.1140/epjc/s10052-016-4271-x}{\emph{Eur. Phys. J. C}
  {\bfseries 76} (2016) 434},
  [\href{https://arxiv.org/abs/1605.05081}{{\ttfamily 1605.05081}}].

\bibitem{Calibbi:2020jvd}
L.~Calibbi, D.~Redigolo, R.~Ziegler and J.~Zupan, \emph{{Looking forward to
  lepton-flavor-violating ALPs}},
  \href{https://doi.org/10.1007/JHEP09(2021)173}{\emph{JHEP} {\bfseries 09}
  (2021) 173}, [\href{https://arxiv.org/abs/2006.04795}{{\ttfamily
  2006.04795}}].

\bibitem{Jho:2022snj}
Y.~Jho, S.~Knapen and D.~Redigolo, \emph{{Lepton-flavor violating axions at MEG
  II}}, \href{https://doi.org/10.1007/JHEP10(2022)029}{\emph{JHEP} {\bfseries
  10} (2022) 029}, [\href{https://arxiv.org/abs/2203.11222}{{\ttfamily
  2203.11222}}].

\bibitem{Knapen:2023zgi}
S.~Knapen, K.~Langhoff, T.~Opferkuch and D.~Redigolo, \emph{{A Robust Search
  for Lepton Flavour Violating Axions at Mu3e}},
  \href{https://arxiv.org/abs/2311.17915}{{\ttfamily 2311.17915}}.

\bibitem{Atwood:1994vm}
D.~Atwood, A.~Kagan and T.~G. Rizzo, \emph{{Constraining anomalous top quark
  couplings at the Tevatron}},
  \href{https://doi.org/10.1103/PhysRevD.52.6264}{\emph{Phys. Rev. D}
  {\bfseries 52} (1995) 6264--6270},
  [\href{https://arxiv.org/abs/hep-ph/9407408}{{\ttfamily hep-ph/9407408}}].

\bibitem{Haberl:1995ek}
P.~Haberl, O.~Nachtmann and A.~Wilch, \emph{{Top production in hadron hadron
  collisions and anomalous top - gluon couplings}},
  \href{https://doi.org/10.1103/PhysRevD.53.4875}{\emph{Phys. Rev. D}
  {\bfseries 53} (1996) 4875--4885},
  [\href{https://arxiv.org/abs/hep-ph/9505409}{{\ttfamily hep-ph/9505409}}].

\bibitem{Cheung:1995nt}
K.-m. Cheung, \emph{{Probing the chromoelectric and chromomagnetic dipole
  moments of the top quark at hadronic colliders}},
  \href{https://doi.org/10.1103/PhysRevD.53.3604}{\emph{Phys. Rev. D}
  {\bfseries 53} (1996) 3604--3615},
  [\href{https://arxiv.org/abs/hep-ph/9511260}{{\ttfamily hep-ph/9511260}}].

\bibitem{CMS:2019kzp}
{\scshape CMS} collaboration, A.~M. Sirunyan et~al., \emph{{Measurement of the
  top quark forward-backward production asymmetry and the anomalous
  chromoelectric and chromomagnetic moments in pp collisions at $ \sqrt{s} $ =
  13 TeV}}, \href{https://doi.org/10.1007/JHEP06(2020)146}{\emph{JHEP}
  {\bfseries 06} (2020) 146},
  [\href{https://arxiv.org/abs/1912.09540}{{\ttfamily 1912.09540}}].

\bibitem{CMS:2019nrx}
{\scshape CMS} collaboration, A.~M. Sirunyan et~al., \emph{{Measurement of the
  top quark polarization and $\mathrm{t\bar{t}}$ spin correlations using
  dilepton final states in proton-proton collisions at $\sqrt{s} =$ 13 TeV}},
  \href{https://doi.org/10.1103/PhysRevD.100.072002}{\emph{Phys. Rev. D}
  {\bfseries 100} (2019) 072002},
  [\href{https://arxiv.org/abs/1907.03729}{{\ttfamily 1907.03729}}].

\bibitem{Galda:2021hbr}
A.~M. Galda, M.~Neubert and S.~Renner, \emph{{ALP \textemdash{} SMEFT
  interference}}, \href{https://doi.org/10.1007/JHEP06(2021)135}{\emph{JHEP}
  {\bfseries 06} (2021) 135},
  [\href{https://arxiv.org/abs/2105.01078}{{\ttfamily 2105.01078}}].

\end{thebibliography}\endgroup

\end{document}